\documentclass[pre,showpacs,reprint]{revtex4-1}

\usepackage{graphicx,amssymb,amsmath,hyperref}
\usepackage{xcolor}
\graphicspath{{figures/}}

\newcommand{\Add}[1]{{\color{red}#1}}

\begin{document}

\title{Intermittency and emergence of coherent structures in wave turbulence of a vibrating plate}
\author{Nicolas Mordant}
\email[]{nicolas.mordant@univ-grenoble-alpes.fr}
\affiliation{Laboratoire des Ecoulements G\'eophysiques et Industriels, Universit\'e Grenoble Alpes \& CNRS}
\author{Benjamin Miquel}
\affiliation{Department of Applied Mathematics, University of Colorado, Boulder, CO 80309-0526, USA.}


\begin{abstract}
We report numerical investigations of wave turbulence in a vibrating plate. The possibility to implement advanced measurement techniques and long time numerical simulations makes this system extremely valuable for wave turbulence studies. The purely 2D character of dynamics of the elastic plate makes it much simpler to handle compared to much more complex 3D physical systems that are typical of geo- and astrophysical issues (ocean surface or internal waves, magnetized plasmas or strongly rotating and/or stratified flows). When the forcing is small the observed wave turbulence is consistent with the predictions of the Weak Turbulent Theory. Here we focus on the case of stronger forcing for which coherent structures can be observed. These structures look similar to the folds and D-cones that are commonly observed for strongly deformed static thin elastic sheets (crumpled paper) except that they evolve dynamically in our forced system. We describe their evolution and show that their emergence is associated with statistical intermittency (lack of self similarity) of strongly nonlinear wave turbulence. This behavior is reminiscent of intermittency in Navier-Stokes turbulence. Experimental data show hints of the weak to strong turbulence transition. However, due to technical limitations and dissipation, the strong nonlinear regime remains out of reach of experiments and therefore has been explored numerically.
\end{abstract}

\maketitle

\section{Introduction}

Historical investigations of wave turbulence were motivated by the quest for a statistical modeling of the sea surface or turbulence in magnetized plasmas~\cite{R9,R1}. It led to the development of the Weak Turbulence Theory (WTT): a statistical theory of wave turbulence in the limits of asymptotically large systems and vanishingly small nonlinearity~\cite{R1,R2,R3}. In this theoretical framework, it is possible to demonstrate the occurrence of an energy cascade in forced, out of equilibrium, dispersive wave systems. This result is remarkable in comparison with the theoretical developments of a statistical theory of turbulence where such an energy cascade is accepted for a century but its justification relies mostly on phenomenological arguments~\cite{Frisch}. 

Although the theory exists for about half a century, clear experimental and numerical evidence of weak turbulence in laboratory experiments is much more recent~\cite{R2,R3,falconrev,R10,R23,HB,Sharon,R8,Aubourg}. Part of the difficulties come from the two above mentioned hypotheses underlying the theory which are difficult to fulfill in the laboratory where the systems are of finite size and dissipative. Dissipation must be overcome by strong enough a forcing in order to develop a non linear stage and this may not be compatible with the hypothesis of weak non linearity. Another difficulty lies in the measurements. Indeed waves are characterized by their propagation. Probing the presence of waves in the non linear system and discriminating them from other structures demands a space and time resolved measurement of the turbulent field~\cite{Mordant,R10,DeikeJFM15,Sharon}. This is usually challenging due to the very nature of turbulence: a wide range of time and space scales. This is particularly difficult in 3D systems. Note that truly weakly non linear numerical simulations are also challenging due to the scale separation between the linear wave period and the much longer nonlinear timescales. This separation makes numerical simulations extremely costly.


In that respect, the vibrating elastic plate has been shown to be a fruitful physical model for studies of wave turbulence. Thin elastic plates can support the propagation of flexion waves. These waves become non linear when the in-plane stretching due to large amplitude deformation becomes significant. As the non linear interactions grow, a turbulent state can develop~\cite{R18,Mordant,R19,Boudaoud,R21}. This state is actually responsible for the noise radiated from a shaken plate and has been utilized for centuries to mimic thunder in theaters~\cite{R17}. The intrinsic 2D character of thin plates made possible experimental studies by means of a space and time resolved profilometry technique as well as numerical simulations either of idealized spatially periodic configurations~\cite{R23,Yokoyama1,Yokoyama2,Yokoyama3} or more realistic ones~\cite{Ducceschi}. These studies investigated stationary out of equilibrium cases and pointed out the consistency with the description offered by the WTT as well as limitations in the applicability of this theory, the most serious of which being due to dissipation in experimental systems~\cite{R23,Humbert,Miquel3}. Non stationary situations either of turbulence decay or turbulence buildup have also been studied that are also consistent with the WTT~\cite{Humbertphd,R24,Auliel}. Furthermore, when the forcing is increased, the vibrating plate displays a transition to a regime of stronger non linearity characterized by the generation of strongly non linear coherent structures that induce deviations from the WTT in terms of power spectra as well as the emergence of intermittency~\cite{Yokoyama1,Yokoyama2,Yokoyama3,R25,Chibbaro}.
In the present article we investigate  both experimentally and numerically this strong forcing regime in much more details. We show that the occurrence of singular structures changes the power spectrum with an additional component on top of the propagating waves. We quantify the observed intermittency and show a very good agreement between experiments and numerical simulations.

The article is organized as follow: in section 2, we introduce the governing equations of thin plates and we give a description of the experimental setup and of the numerical methods. The different regimes potentially encountered as the forcing intensifies are presented and qualitatively described in section 3. We identify in section 4 the space-time spectral signature of the coherent structures. From this signature, we infer a first estimate of their mean velocity. A dual analysis in physical space is carried out in section 5, which confirms the physical picture of coherent singularities propagating following a velocity distribution that we relate to the space-time spectrum of the motion. Finally, we demonstrate in section 6 that the emergence of singular structures is associated with the apparition of intermittency in the system. The last section is dedicated to concluding remarks.

\section{Methods}

\subsection{The wave equation}

The dynamics of the deformation $\zeta$ of a shaken thin elastic plate (flat at rest) follows the F\"oppl-von Karman equations:\begin{subequations}
\begin{align}
\partial _{tt} \zeta &= -\frac{E h^2}{12\rho(1-\sigma^2)}\Delta^2\zeta + \frac{1}{\rho}\left\{\zeta,\chi \right\}+\mathcal F+\mathcal D\, ,\label{eq_FVK1}\\
\Delta^2 \chi& =-E\frac{\left\{\zeta, \zeta \right\}}{2} \,\, ,\label{eq_FVK2}
\end{align}
\label{eq_FVK}
\end{subequations}
where the physical properties of the material are described by the following coefficients: Young's modulus $E$, Poisson's ratio $\sigma$, the density $\rho$. $\mathcal F$ and $\mathcal D$ are forcing and dissipation respectively. The brackets $\left\{\cdot,\cdot\right\}$ denote the bilinear differential operator 
\begin{equation}
\left\{\zeta,\chi\right\}=\partial_{xx}\zeta\partial_{yy}\chi + \partial_{yy}\zeta\partial_{xx}\chi -2 \partial_{xy}\zeta\partial_{xy} \chi\,.
\end{equation}
In the limit of vanishingly small amplitude deformations, pure flexion yields the linear part of equation (\ref{eq_FVK1}) which translates into the flexion waves dispersion relation: $\omega=\pm ck^2$ with $c=\sqrt{\frac{E h^2}{12 \rho(1-\sigma^2)}}$. 

As the amplitude of the deformation grows, in-plane stretching comes into play and generates non linear terms in the FvK equations. Despite being in the limit of linear elasticity for the bulk material, non linear terms arise for geometrical reasons: the local stretching induced in an initially flat surface is proportional to the Gaussian curvature $G$ of the deformation $\zeta$ whose full expression and leading order in the limit of a weak slope are respectively:
\begin{subequations}
\begin{align}
G(\zeta)&=\frac{\partial_{xx}\zeta\partial_{yy}\zeta - (\partial_{xy}\zeta)^2}{(1+(\partial_x \zeta)^2+(\partial_y \zeta)^2)^2}\\
& = G_0(\zeta) \left[1+\mathcal{O}\left(\left|\boldsymbol{\nabla} \zeta\right|^2 \right) \right] \quad\mathrm{where}\quad G_0(\zeta) = \frac{\left\{\zeta, \zeta \right\}} {2}
\end{align}
\end{subequations}
 Formally, in-plane stretching is conveniently taken into account at the lowest order by introducing the auxiliary Airy function $\chi$. The Airy function obeys equation (\ref{eq_FVK2}) where the leading order of the Gaussian curvature $G_0$ acts as a source term. For this reason, the statistical properties of the Gaussian curvature are detailed in sections 5 and 6 as the wave turbulence intensity increases in a vibrating plate.

\subsection{Experimental setup}

The description of the experiment is briefly recalled here and the reader is referred to earlier publications for more details (see \cite{R19,R21}). A $1\times2$~m$^2$ stainless steel plate of thickness 0.4~mm is hanging vertically under its own weight. The plate is set in motion by an electromagnetic shaker oscillating at 30~Hz anchored in the lower part of the plate. The vibration is recorded by a Fourier transform profilometry technique \cite{R22,Cobelli1} that provides the deformation of the plate resolved both in time (at frequencies up to a 5 kHz) and space (over about a square meter). Seven experiments (run A to G) have been realized and the relative range of injected power $P$ is shown in table~\ref{param}. The maximum power is 50 times the smallest. For stainless steel, $E\approx 2\, 10^{11}$~Pa, $\sigma\approx 0.3$, $\rho=7800$~kg/m$^3$ thus $c\approx 0.61$~m$^2$/s.

\begin{table}
\centering    
\begin{tabular}{c|ccccccc}
exp. run & A & B & C & D & E & F & G \\
\hline
$P$ (relative) & 1& 4 & 9 &16 & 25 & 36 & 50
\end{tabular}

\begin{tabular}{c|cccccccc}
num. run & 1 & 2 & 3& 4 & 5 & 6 & 7 & 8\\
\hline
$P$ (relative) & 1 &3.5 &14 &60 &245 &910 & 3760&11100
\end{tabular}
\caption{Parameters of the experiments (top) and numerical simulations (bottom)
\label{param}}
\end{table}

\subsection{Numerical simulations}

The numerical simulation has been described in \cite{R23}. The forced F\"oppl-Von Karman equations of a square plate with periodic boundary conditions are time-stepped using a pseudo-spectral scheme. The linear dissipation $\mathcal D_k=-\gamma_k \partial_t\zeta_k$ can be chosen either similar to the experimentally measured one or as an ``ideal'' dissipation that is zero below a cutoff wavenumber (chosen here as $k_c/2\pi=100$~m$^{-1}$). The former case has been shown to reproduce qualitatively the measurements~\cite{R23} whereas the latter case reproduces the scaling properties of the theoretical predictions from the WTT. In this article we consider only the second sort of ideal dissipation.

The system is driven out-of-equilibrium by a spectral forcing of the form:
\begin{equation}
\mathcal{F}_{\boldsymbol{k}}(t) = F_0 \mathrm{e}^{\mathrm{i}\psi_{\boldsymbol{k}}}\exp\left[-\frac{\left(\left|\boldsymbol{k}\right|-k_f\right)}{2\sigma_k^2}\right] \cos\left(\omega_{\boldsymbol{k}} t + \phi_{\boldsymbol{k}} \right)
\end{equation}
which corresponds to forcing with an amplitude $F_0$ an isotropic crown of wavenumbers centered around the value $k_f=5\pi$~m$^{-1}$ with a width $\sigma_k=1$~m$^{-1}$. Each mode is forced to its linear frequency $\omega_{\boldsymbol{k}}$ and the random phases $\psi_{\boldsymbol{k}}$ and $\phi_{\boldsymbol{k}}$ ensure a disordered forcing.  In the eight runs (named run 1 to 8) reported in this paper, we kept the parameters of the equation constant such that to correspond to a $2\times 2$ m$^2$ plate of steel of thickness $0.4$ mm. The magnitude of the forcing $F_0$ has been increased by factors of 2 (i.e. the relative amplitude spans 1 to 128). The corresponding injected powers (proportional to $F_0^2$) span four orders of magnitude (table~\ref{param}) which is much larger than what is possible in the experiments. 

\section{Morphological evolution of the deformation of the vibrating plate with the forcing intensity}


\begin{figure}[!htb]
\centering
\includegraphics[width=8cm]{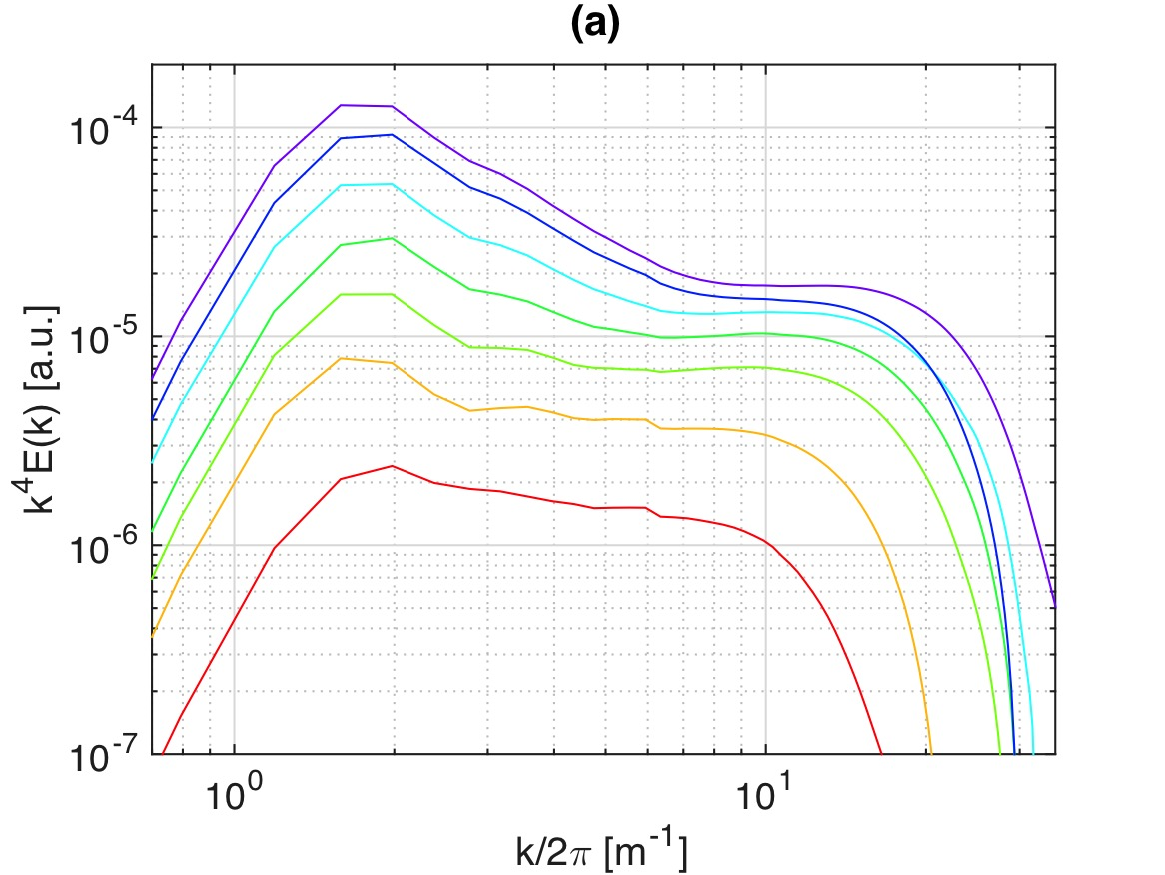}
\includegraphics[width=8cm]{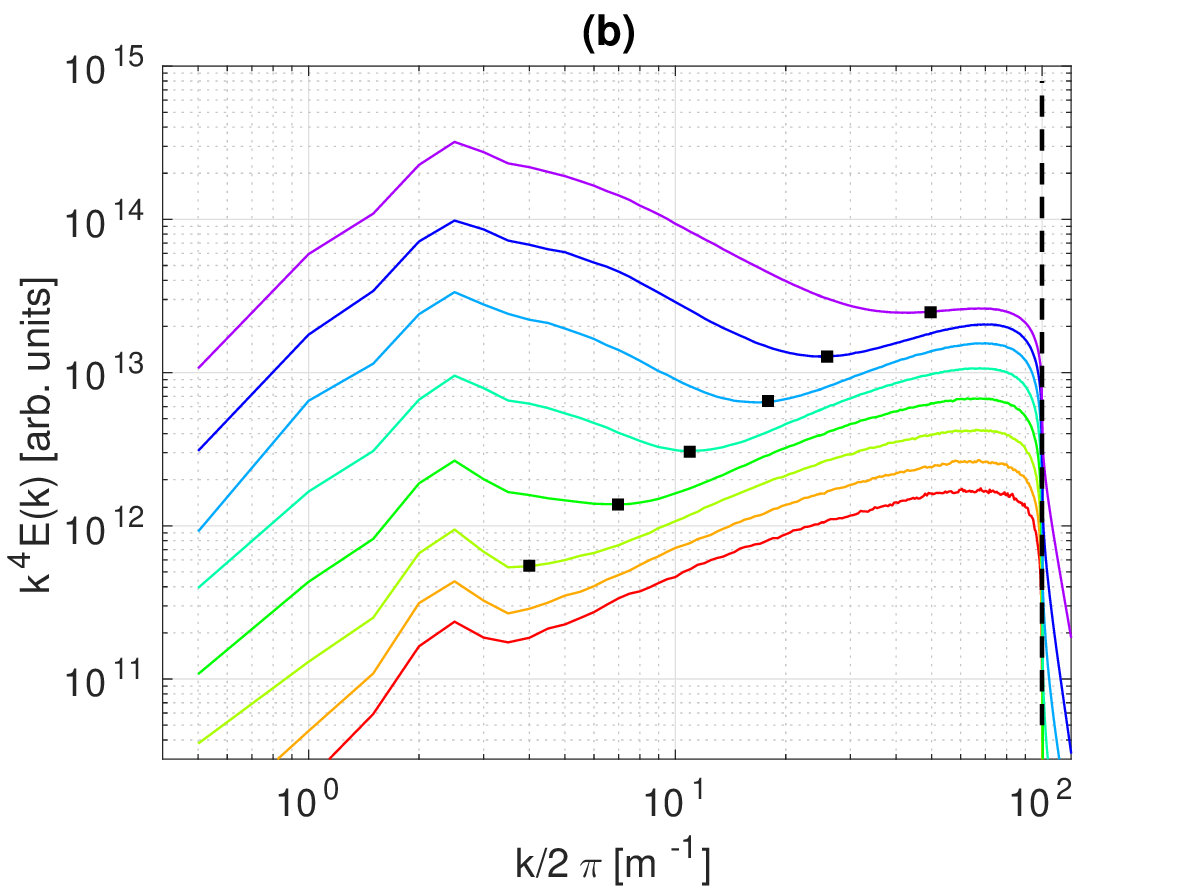}
\caption{Evolution of the Fourier power spectrum $E^{\eta}(k)$ of the deformation $\eta$ of the plate (integrated over the directions of the wavevector $\mathbf k$) as a function of the forcing intensity. (a) experiment. The average forcing power is $P=[1,4,9,16,25,36,50]$ (arb. unit) from bottom to top. The power is proportional to the square of the force magnitude (at a given excitation frequency). (b) numerical simulations. The input power is $P=[1,3.5,14,60,245,910, 3760,11100]$ (arb. unit) from bottom to top and thus evolves on a much larger range than the experiment. The dissipation is also quite different (see text). The vertical dashed line corresponds to the wavenumber over which the dissipation operates. The forcing operates at $k/2\pi=2.5$~m$^{-1}$ (that corresponds to 30~Hz for linear waves as in the experiment). Black squares show the change of behavior of the spectrum between structure dominated regime (at low $k$) and weak turbulence at high $k$.}
\label{sp}
\end{figure}

\begin{figure*}[!htb]
\centering
\includegraphics[width=18cm]{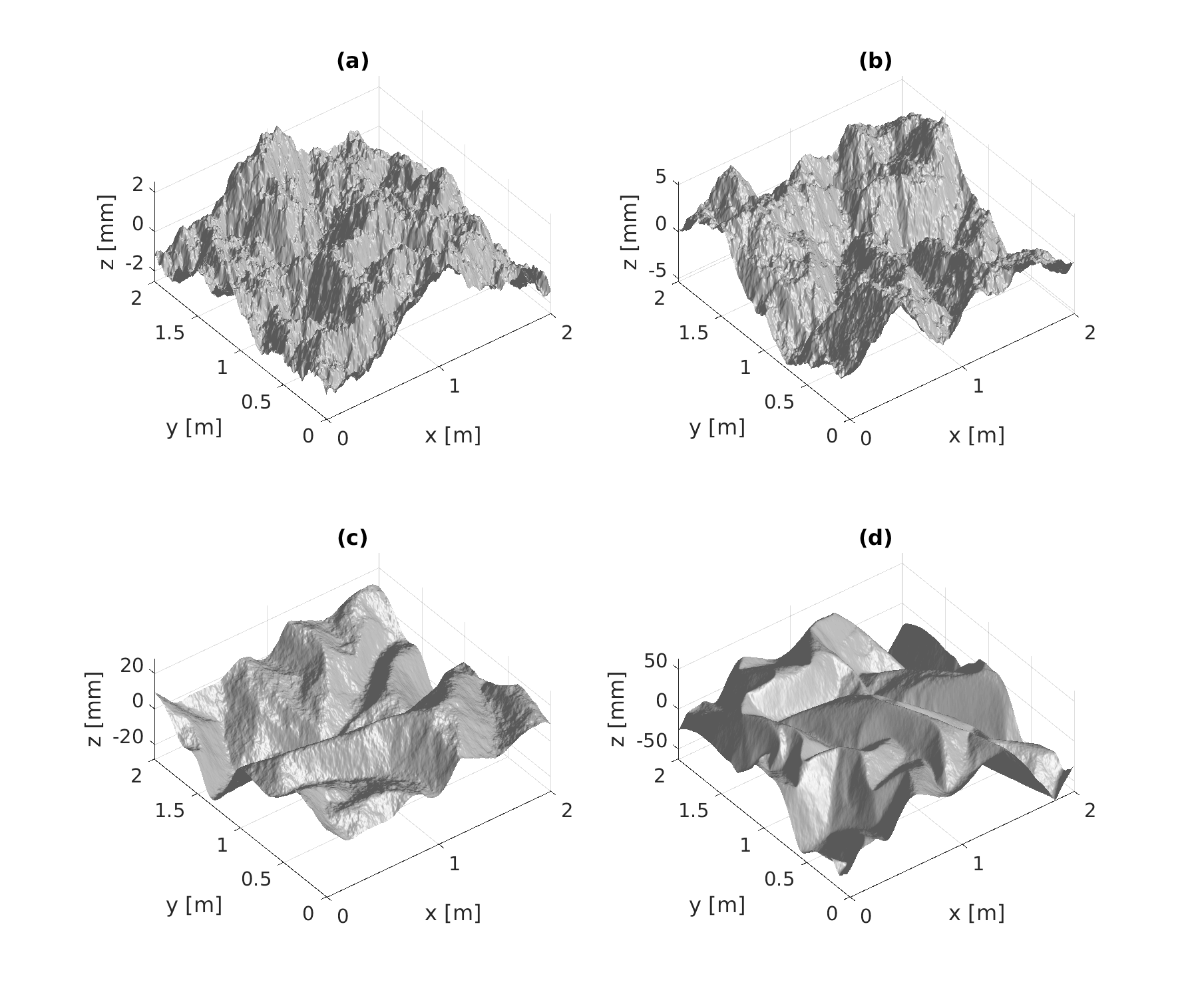}
\caption{Snapshots of the deformation of the plate for various forcing intensities (numerical simulation). The corresponding run number is displayed in the title. The input power is increasing with the run number and the relative values $(1,14,245,3760)$ from (a) to (d) respectively.}
\label{ex}
\end{figure*}
We characterize in this section the different dynamical regimes observed as the forcing intensity increases.
The shape of the Fourier power spectrum $E^{\eta}(k)$ of the deformation $\eta$ (integrated over the direction of the wavevector $\mathbf k$, assuming isotropy) has been previously observed to evolve with the forcing intensity both experimentally and numerically~\cite{R25,Yokoyama1,Yokoyama2,Yokoyama3}. \Add{We illustrate these previously published observations with a new set of numerical simulations with a higher forcing magnitude}. As is seen in fig.~\ref{sp}, the low wavenumber part of the spectrum steepens progressively as the forcing increases. The same trend is visible on both experiment and numerical simulation. In comparison to the experiment, the numerical simulation is forced much more strongly. Qualitatively the strongest experiment (run G) appears similar to the numerical run 4 as far as the low wavenumber part is concerned. A direct quantitative comparison is not possible as the dissipation processes are very different as well as the boundary conditions. These observations have been related to the creation of strongly nonlinear structures that can be observed in physical space~\cite{R25}. Indeed fig.~\ref{ex} shows four snapshots of the deformation of a vibrating plate at four increasing values of the forcing amplitude. The spatial structure of the rugosity of the deformation is obviously changing when the forcing is increased. Large scale structures incorporating ridges and developable cones~\cite{RMPWitten} are getting more and more visible when the amplitude of the forcing grows. At the highest forcing intensity the deformation resembles crumpled paper (be aware though that the magnitude of the vertical deformation has been strongly amplified in fig.~\ref{ex}). These structures can also be identified in spectral space: the emergence of the steeper part of the spectrum at low wavenumber has been shown to correspond to the emergence of ridges and cones~\cite{R25}.  

\begin{figure}[!htb]
\centering
\includegraphics[width=8cm]{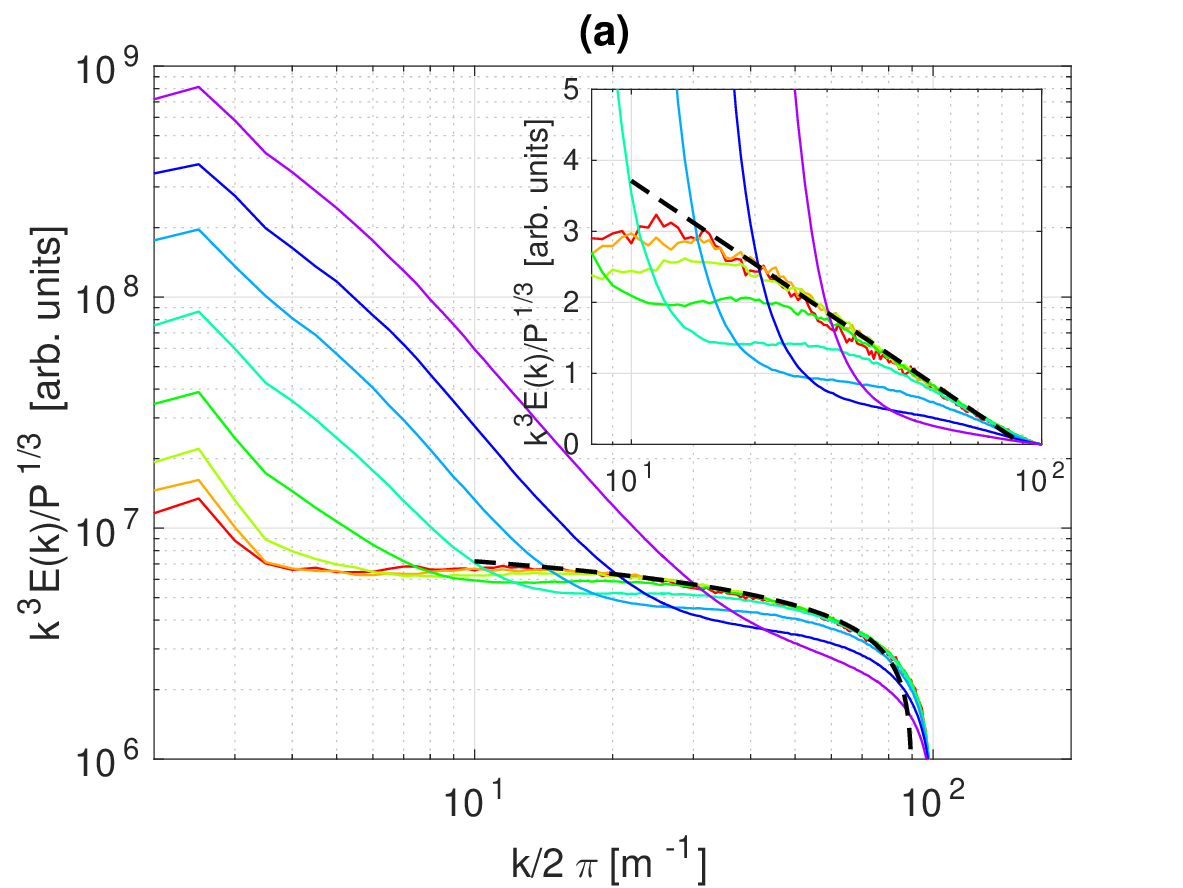}
\includegraphics[width=8cm]{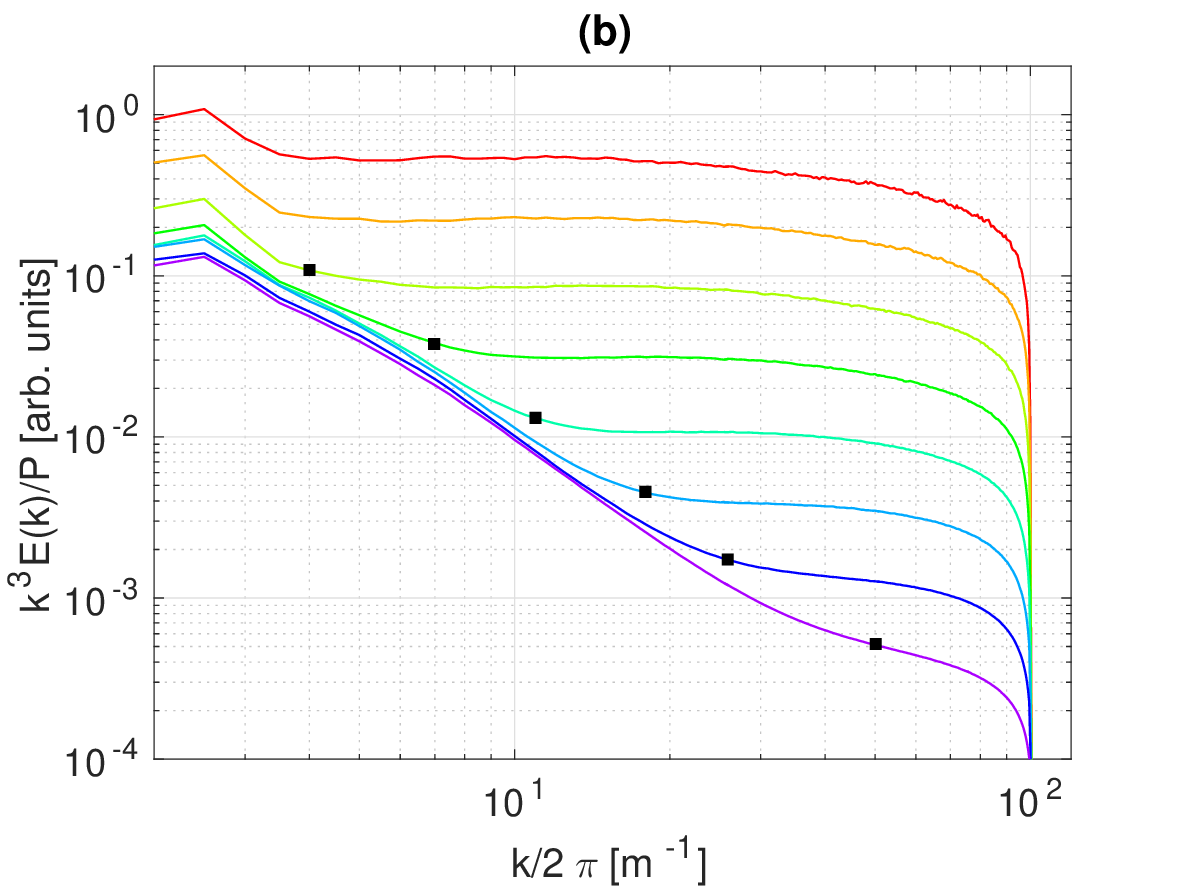}
\caption{Scaling properties of the spectrum of the deformation. The color code is the same as in fig.~\ref{sp}. (a) spectra normalized by $k^3/P^{1/3}$ in order to compare to the prediction of the WTT that predicts a stationary spectrum $E^{\eta}(k)\propto P^{1/3}\frac{\log^{1/3}(k^*/k)}{k^3}$. The dashed line correspond to a fitted $\log^{1/3}(k^*/k)$ with $k^*/2\pi=90$~m$^{-1}$. The insert is a zoom of the logarithmic region in semilogarithmic scale. (b) Spectrum normalized by $k^3/P$. This normalisation was chosen to illustrate the linear scaling with $P$ of the low frequency part of the spectrum. Black squares show the transition as in fig.~\ref{sp}.}
\label{spsc}
\end{figure}

\begin{figure}
\begin{center}
\includegraphics[width = 0.5\textwidth]{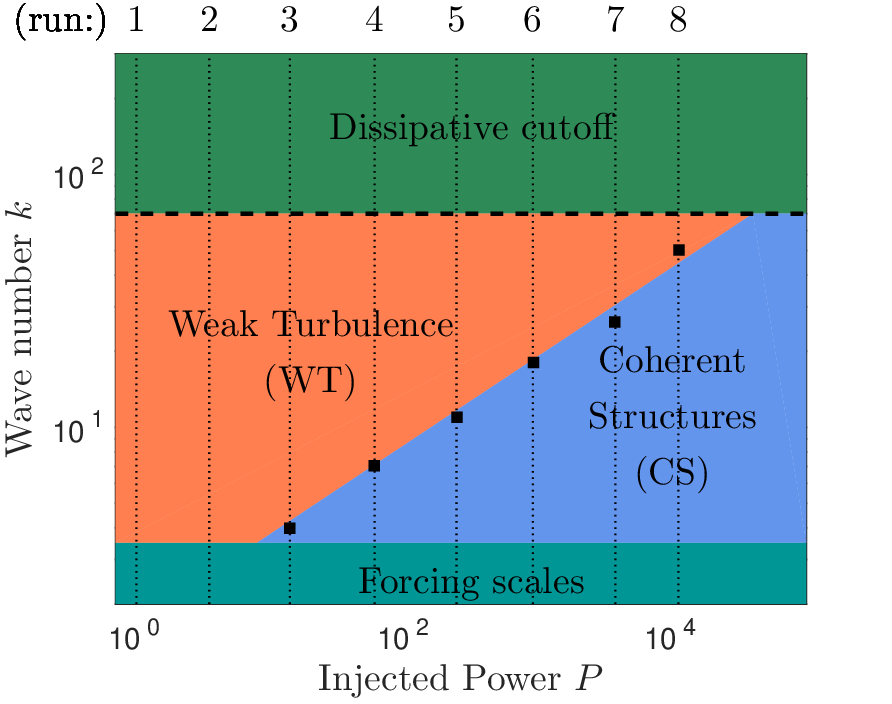}
\caption{\label{schema_regimes}Sketch of the different regimes encountered in the spectrum of the motion as the injected power increases. The different numerical runs labeled from 1 to 8 (see table~\ref{param}) are  represented by vertical dotted lines. The largest scales and the smallest scales are dominated by the forcing and the dissipation, respectively. The intermediate scales exhibit a Weak Turbulence (WT) scaling for weak forcing, or are dominated by coherent structures (CS) for very large forcing. At intermediate forcing, the large scales are populated by the coherent structure while the small scales follow a WT dynamics. The transition observed in the spectra represented in figure~\ref{sp} and \ref{spsc} is denoted by plain squares. They correspond to the change from the the strongly nonlinear regime ($E^{\eta}(k)\propto Pk^{-5}$) to the KZ spectrum ($E^{\eta}(k)\propto P^{1/3}\frac{\log^{1/3}(k^*/k)}{k^3}$). Thus the transition wavenumber $k_c$ is expected to follow roughly $k_c\propto P^{1/3}$, discarding logarithmic corrections.}
\end{center}
\end{figure}

We illustrate this phenomenon in figure~\ref{spsc}(a) where we display the deformation spectra obtained numerically in runs 1-8 and normalized by $k^3/P^{1/3}$. This scaling corresponds to the WTT prediction which derives a stationary spectrum of the form:
\begin{equation}
E^{\eta}(k)\propto P^{1/3}\frac{\log^{1/3}(k^*/k)}{k^3}\, .
\end{equation} 
Observe that in absence of coherent structures (weakest runs 1 and 2) the wave turbulence scaling is observed as the spectra collapse on top of each other. They also follow quite convincingly the logarithmic trend predicted by WTT. At intermediate forcing and for low wavenumbers, the spectra departs from the weak case with a large amplitude steeper part. It means that the low $k$ part of the spectrum follows a distinct scaling both in $k$ and $P$. At higher wavenumber these spectra recover the weak amplitude logarithmic scaling from below and the $P^{1/3}$ scaling. For stronger forcing where coherent structures are observed (runs 5-7), three different regions are identified successively in the spectra as frequency increases: the low frequency part is heavily affected by the apparition of ridges and cones and hence develops further a scaling dubbed ``Coherent Structures'' (CS) scaling which is distinct from the $P^{1/3}$ WTT scaling; the small scales remain potentially in a weak wave turbulence regime, as the high frequencies still follow the weak scaling; the highest frequencies at the end of the cascades exhibit a dissipation dominated dynamics associated with a viscous cutoff of the spectrum. This transition in the regimes observed in the spectrum are summarized schematically in figure~\ref{schema_regimes}. Note that the existence of the intermediate regime of weak wave turbulence depends on the relative intensities of the forcing and the dissipation: we observed with our most vigorous forcing (run8) a direct transition from the CS scaling to the viscous cutoff and a complete absence of the WTT scaling (see upper curve of figure~\ref{spsc}). 
Finally, we identify the low frequency CS scaling with $P$ to be proportional to $P$ as demonstrated by displaying the deformation spectra normalized by $k^3/P$ on fig.~\ref{spsc}(b). \Add{The spectrum of the singularities has been predicted in~\cite{Boudaoud} to be proportional to $1/k^{-5}$ and observed to be proportional to $P$ (thus a global scaling $P/k^5$). This scaling has been first observed in~\cite{R25} and is confirmed in fig.~\ref{spsc}(b). The KZ  spectrum is scaling as $P^{1/3}/k^3$. The former dominates at low $k$ whereas the latter dominates at large $k$. The two spectra are equal at a critical wavenumber $k_c$ such that $P^{1/3}/k_c^3\approx P/k_c^5$ which corresponds to $k_c\sim P^{1/3}$. This behavior is confirmed in fig.~\ref{schema_regimes} where the boundary of the WT and CS regimes scales as $P^{1/3}$.}


\section{Spatio-temporal spectral signature of coherent structures}

\begin{figure*}[!htbp]
\centering
\includegraphics[width=\textwidth]{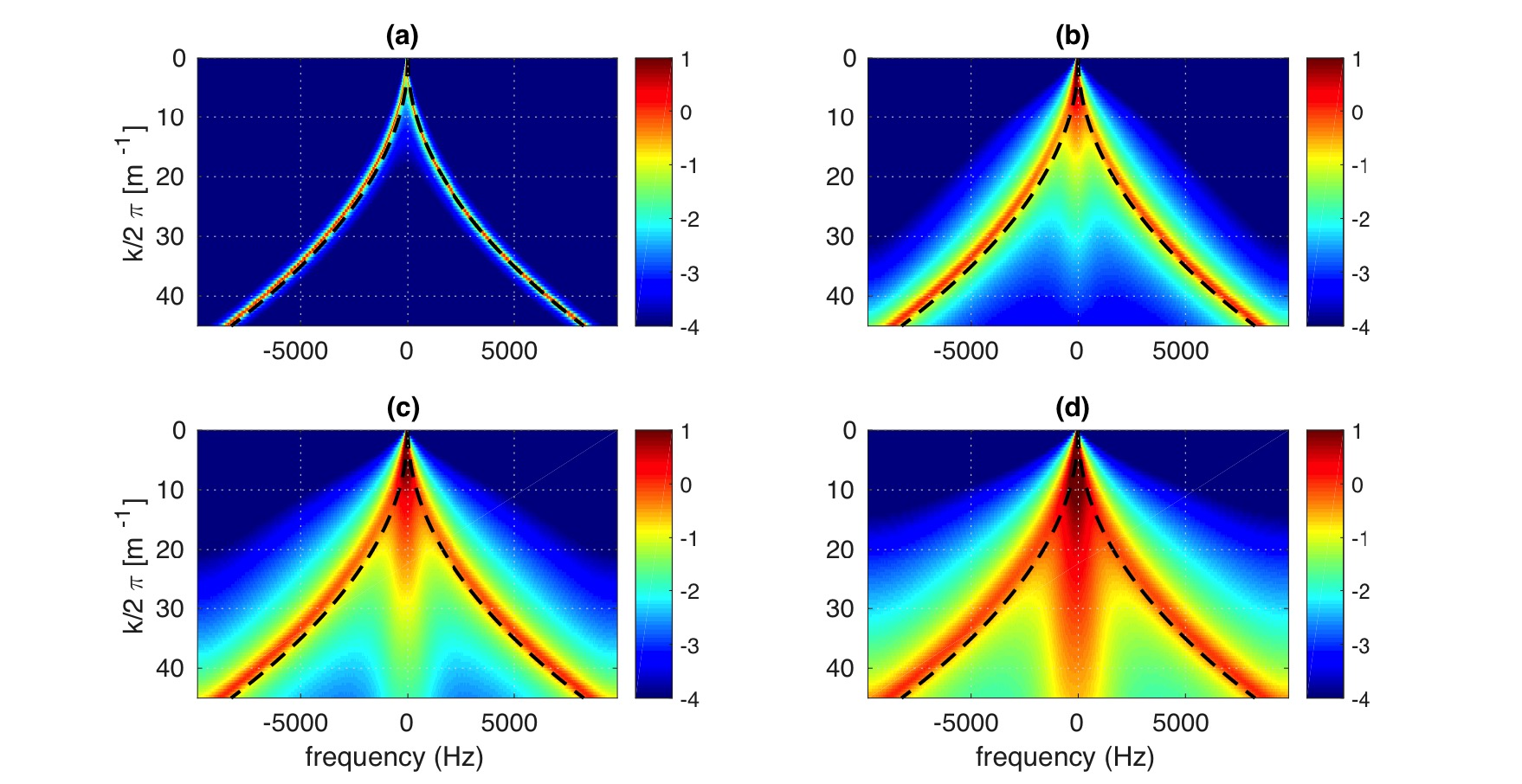}
\caption{Examples of the spectrum $E^{\eta}(k,\omega)$ obtained from the numerical simulations, for run1 (weakest forcing), run6, run7 and run8 (strongest) ($P=1$, 910, 3760 \& 11100, from (a) to (d) respectively). For a better contrast the displayed quantity is actually the normalized spectrum $k^4E^{\eta}(k,\omega)$.}
\label{stfs}
\end{figure*}
The natural way to investigate the dynamics of propagating waves is the space-time energy spectral density of the wave elevation $E^{\eta}(\boldsymbol{k},\omega)$, which we use in this section to obtain a more subtle characterization of the structures presented above. $E^{\eta}(\boldsymbol{k},\omega)$ is obtained by computing the Fourier transform of the plate deformation $\eta(x,y,t)$ both in time and space so that to obtain $\eta(\mathbf k,\omega)$ and thus $E^{\eta}(\boldsymbol{k},\omega)=\langle |\eta(\mathbf k,\omega)|^2\rangle $. In order to get a 2D picture that is easier to represent and comprehend than a full 3D spectrum, one takes advantage of the isotropy of the system by computing $E^{\eta}(k,\omega)$ the spectrum summed over the directions of the wave vectors. This spectrum is shown in fig.~\ref{stfs} for a collection of increasing forcing magnitudes in the numerical simulations. The case of weakest forcing is typical of weak turbulence, for which energy is localized in the very vicinity of the linear dispersion relation. A spectacular effect of the increase of the forcing magnitude is the emergence of a ``tongue'' of energy centered around the zero frequency, in addition to the dispersion relation. This tongue is more and more present when increasing the forcing at its highest magnitudes. This extra component of the spectrum is the most obvious trace of the presence of the structures in the dynamics.

\begin{figure}[!htb]
\centering
\includegraphics[width=9cm]{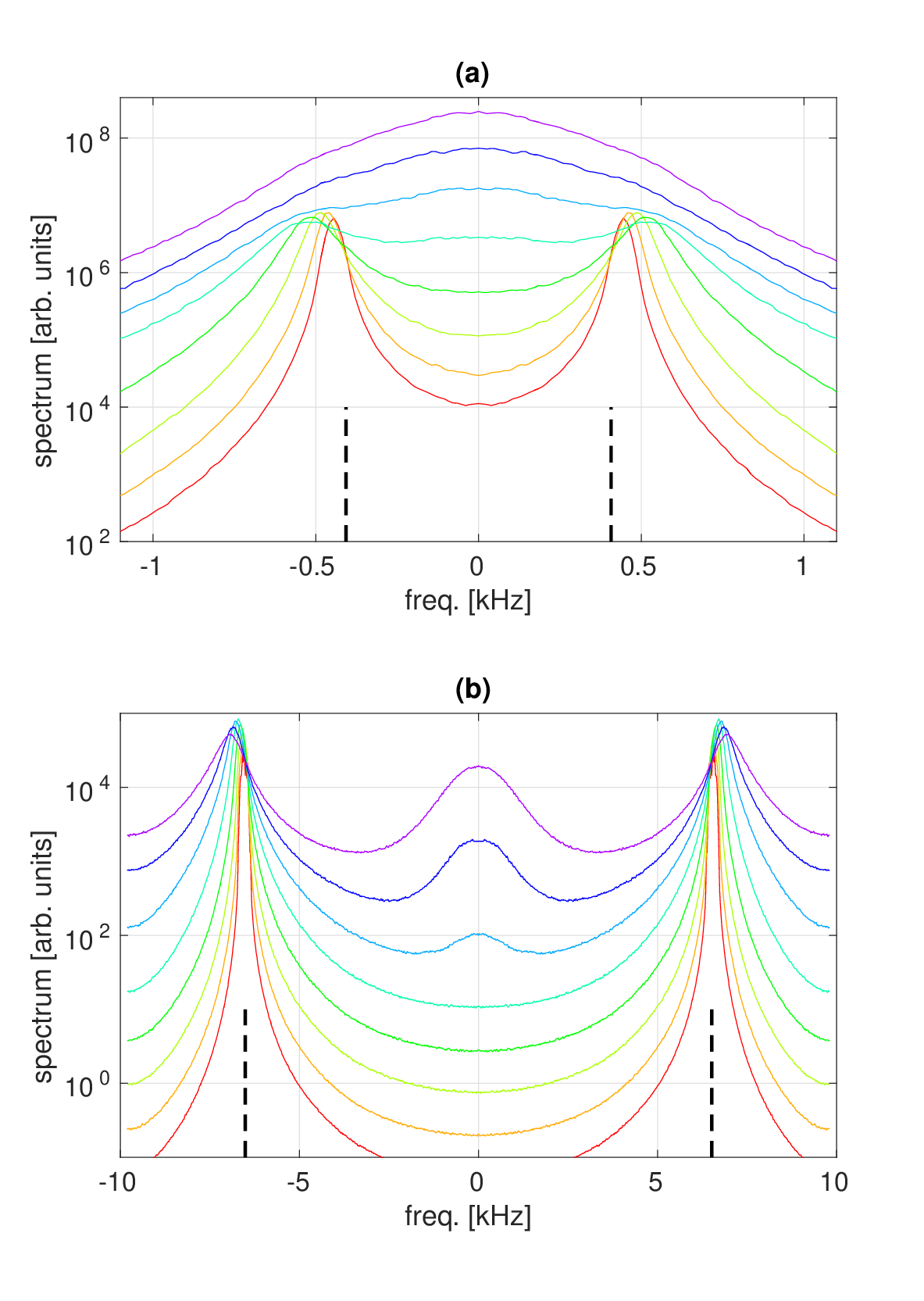}
\caption{Cuts of the $k,\omega$ spectrum for two given values of $k$ ((a) $k=20\pi$ and (b) $k=80\pi$~m$^{-1}$) and for all the values of the forcing magnitude. The linear values of the  linear wave frequency at the given values of $k$ are shown with vertical dashed lines. }
\label{cutsp}
\end{figure}
The growth of the tongue of energy is also visible in fig.~\ref{cutsp} that shows cuts of $E^{\eta}(k,\omega)$ at two given values of the wavenumber $k$.  At low forcing the spectrum is strongly peaked at frequencies close to the linear dispersion relation with the previously mentioned slight non linear shift.
As the non linearity of the waves in the turbulent state is decreasing with the wavenumber, so does the non linear shift: the frequency of the peak is almost equal to the linear frequency at the highest wavenumbers. The widening of the peak with increasing forcing is also visible. At low wavenumber, the width of the peak gets of similar order of magnitude than its central frequency so that at the highest forcing magnitude the peaks are merging as seen for $k/2\pi=10$ m$^{-1}$ (top curves of top panel of fig.~\ref{cutsp}). In the range of wavenumbers for which the peaks have merged, a continuous band of wavenumbers and frequencies can be excited, which actually opens the possibility of observing more complex structures than propagating sine waves. At the highest forcing magnitude, a third peak is observed centered on zero frequency. This peak corresponds to non propagative structures.

\section{Identification of the coherent structures, statistical properties and intermittency}

\subsection{Structures and curvature}

As we mentioned when we introduced the F\"oppl-von K\`arm\`an equations, the source of the non linearity is the Gaussian curvature $G$ that induces stretching of the plate which eventually generates coherent structures. Thus we will focus on the study of the statistical properties of this curvature (or rather its simplified form $\{\zeta,\zeta\}$). In addition to the Gaussian curvature we also study the average curvature 
\begin{subequations}
\begin{align}
L(\zeta)&=\frac{(1+\partial_x\zeta^2)\partial_{xx}\zeta-2\partial_x\zeta\partial_y\zeta\partial_{xy}\zeta+(1+\partial_y\zeta^2)\partial_{yy}\zeta}{(1+\partial_x\zeta^2+\partial_y\zeta^2)^{3/2}}\, ,\\
& = L_0(\zeta) \left[1+\mathcal{O}\left(\left|\boldsymbol{\nabla} \zeta\right|^2 \right) \right] \quad\mathrm{where}\quad L_0(\zeta) = \partial_{xx}\zeta + \partial_{yy}\zeta\, .
\end{align}
\end{subequations}Since $L_0$ is a linear function of $\zeta$, we use it to estimate $L$. The {\it rms} value of the gradient of the deformation is at most 0.36 for run 8 and significantly lower for the other runs. Hence the square gradient is of the order of 10\% which justifies the use of the approximate expressions of $L\approx L_0$ and $G\approx G_0$ in the following.

\begin{figure*}[!htb]
\centering
\includegraphics[width=\textwidth]{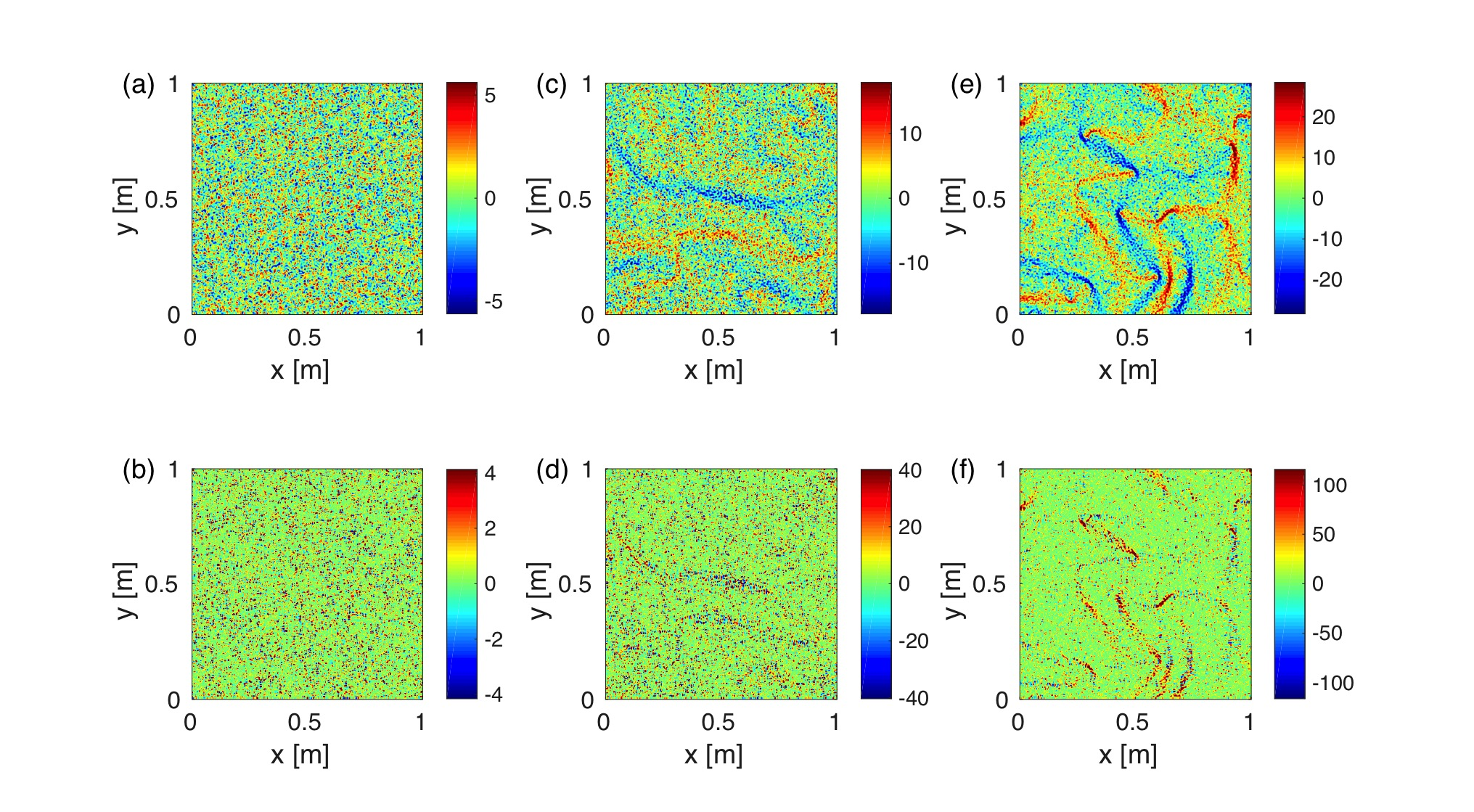}
\caption{Snapshots of both sorts of curvature $L$ and $G$ at various forcing intensities in the DNS. Top line: $L$, bottom line $G$. (a)\&(b): run~1 (weak forcing), (c)\&(d): run~6 (intermediate forcing), (e)\&(f): run~8 (strongest forcing). The input power are respectively $P=1$, 910 and 11100.}
\label{slk}
\end{figure*}
Figure~\ref{slk} displays some snapshots of both sorts of curvature for weak and strong forcing intensities. The curvatures are obviously very small scale quantities as expected from their expression that involve second order derivatives. At the weakest forcing, the deformation field is close to Gaussian statistics and both curvatures look totally disordered. When increasing the forcing a spatial organisation of the curvature becomes obvious. At intermediate forcing an organisation under the form of stripes appear. At the strongest forcing, very intense localized structures dominate over the background. These localized events show a spatial structure with crescent shapes. These crescent shapes have been also observed for $G$ in the case of static developable cones, near the tip of the singularity and they are the footprint of stress focussing in the plate's deformation~\cite{RMPWitten,Cerda,Benamar}.

\begin{figure*}[!htb]
\centering
\includegraphics[width=18cm]{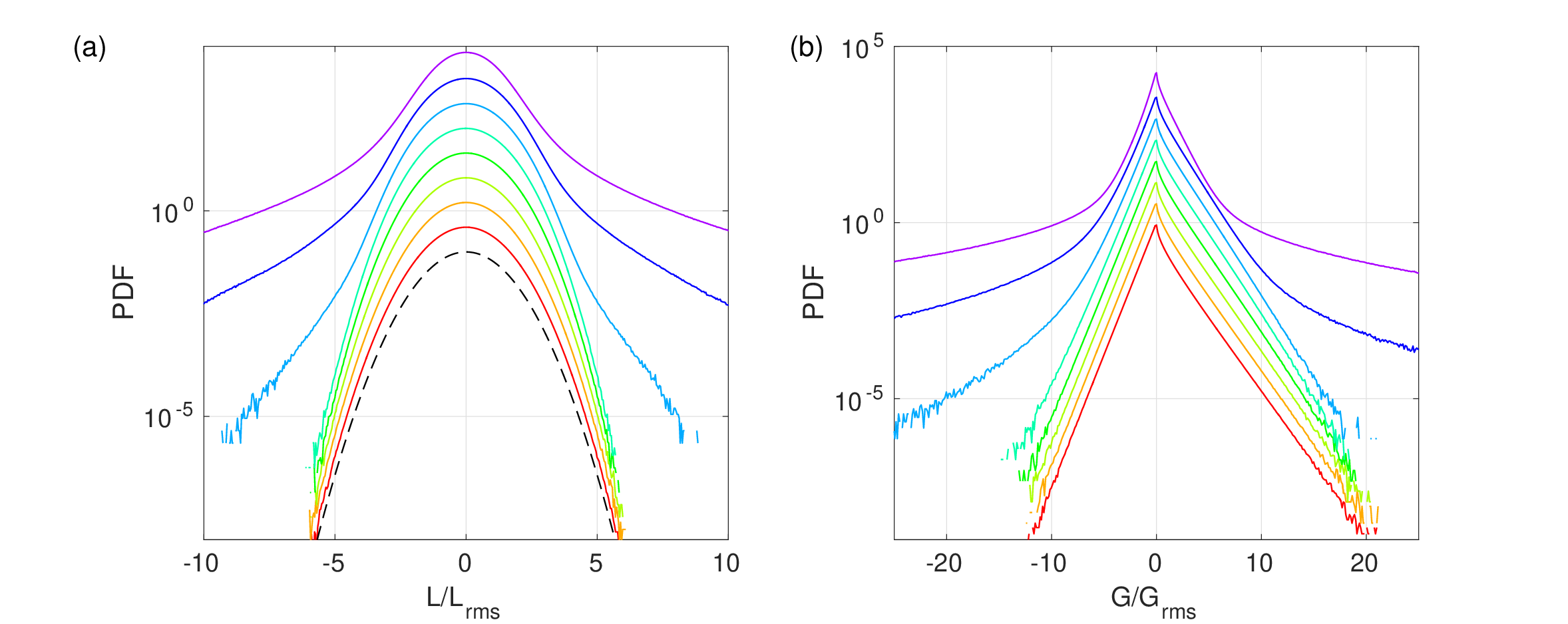}
\caption{PDF of $L_0$ (left) and $G_0$ (right) at various forcing intensities. In each figure: from top to bottom run~1 to run~8 (the curves have been shifted vertically by a factor 4 for clarity). Left: the dashed line is a Gaussian distribution.}
\label{pdflk}
\end{figure*}
This observation is confirmed by the shape of the probability density functions (PDF) shown in fig.~\ref{pdflk}. At weak forcing, the distribution of $L$ is close to Gaussian as expected for weak turbulence. In very weak turbulence, the statistics are expected to show very little departure from Gaussian statistics. The distribution of $G$ is not Gaussian in the weak case which is expected since $G$ is quadratic in the second order derivatives of $\zeta$. The distribution of $G$ is seen to be asymmetric with power law tails, the positive events being more probable than the negative ones. A sharp cusp is also observed for $G$ close to zero. This shape is reminiscent of the distribution of the product of two correlated Gaussian random variables~\cite{Aumaitre} that displays such an asymmetric shape with exponential tails. Here the leading order of the Gaussian curvature $G_0=\partial_{xx}\zeta\partial_{yy}\zeta - (\partial_{xy}\zeta)^2$ is more complex than just the product of two random variables as it is the difference between two quadratic terms, nevertheless the shape of its distribution is clearly related to this simple case. When the intensity of the forcing is increased, the shape deviates from the Gaussian case. The distribution of both curvature displays very wide tails. These tails are symmetric for $L$ as expected from the $\zeta\leftrightarrow-\zeta$ symmetry of the FvK equations and  from $L$ being an odd function of $\zeta$. By contrast the distribution of $G$ is not symmetric: this can be understood by noticing that $G$ is an even function of $\zeta$ so that negative events stem from different relative contributions of the two terms $\partial_{xx}\zeta\partial_{yy}\zeta$ and $\partial_{xy}\zeta ^2 $ instead of being merely obtained by symmetry $\zeta \rightarrow -\zeta$. Extreme negative curvatures (saddle-like structures) are more probable than the positive ones. 

Thus our observations show the appearance of localized intense structures at strong forcing that are consistent with our previous interpretation of these structures being developable cones. At intermediate values of the forcing, a spatial organization of the curvature is also particularly visible in the case of the average curvature $L$ in fig.~\ref{slk}. Strong events are organized spatially in stripes of either strong positive or negative values. This spatial organisation is the trace of the large scale folds that can be seen on the snapshots of the deformation in fig.~\ref{ex}. The traces of the folds are less visible on $G$ as for a simple fold the Gaussian curvature is expected to be zero. The folds are visible nonetheless because of the fluctuations at the smallest scales that are sensitive to the flexion present in the vicinity of the crest of the folds. 

\subsection{Motion of the structures}
\label{PTV}
\begin{figure}[!htb]
\centering
\includegraphics[width=8cm]{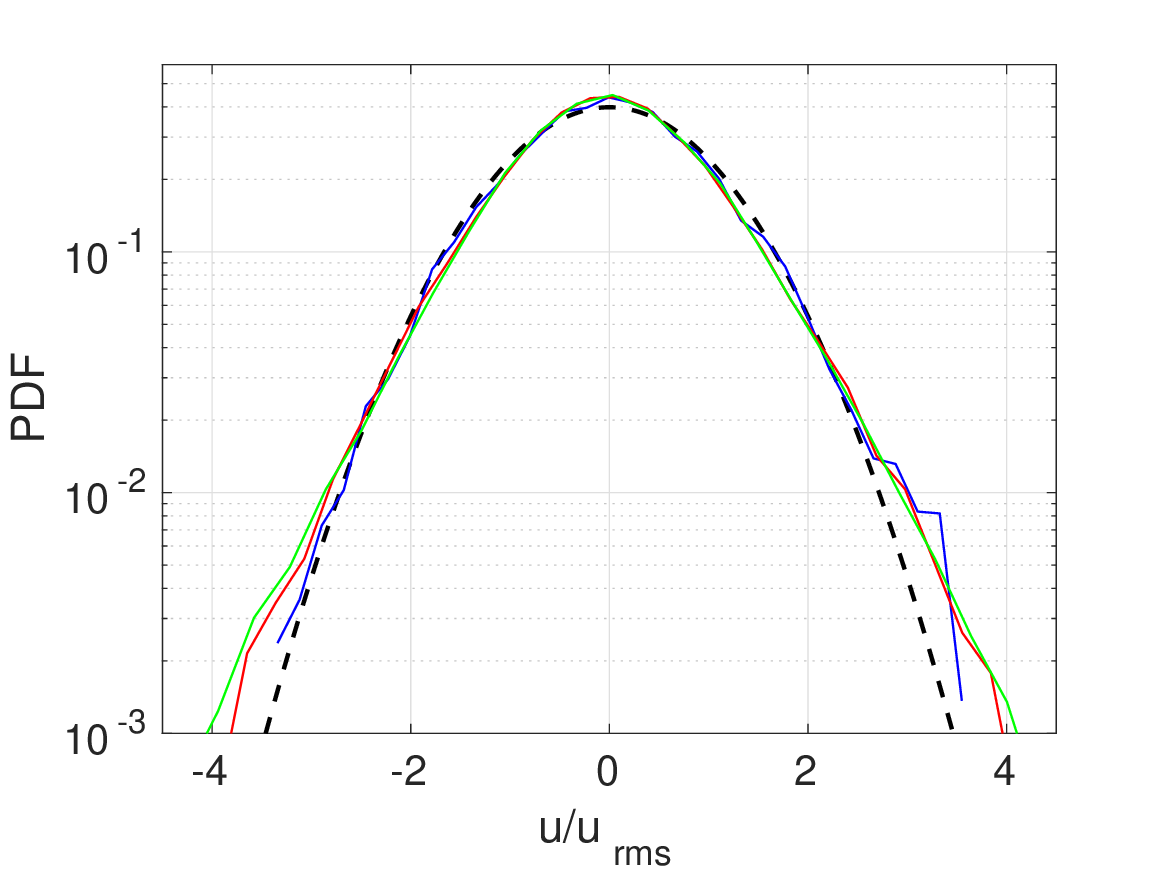}
\caption{Distribution of the velocity of the strong events of $|L|$ at the 3 highest forcing intensities (run 6 to run 8, blue, red and green resp.). The dashed line is a Gaussian distribution. The corresponding forcing powers are respectively $P=910$, 3760 \& 11100.}
\label{pdfpiv}
\end{figure}
We perform a detection and tracking in time of the strong events of curvature. D-cones are characterized by strong values of the Gaussian curvature. One technical issue for detection of D-cones is that the Gaussian curvature is extremely sensitive to the presence of small scale waves that propagate on top of the structures. They induce very strong small scale fluctuations of the Gaussian curvature that oscillate between very large negative and positive values. The detection of the position of the structures is thus very noisy making the tracking of the structures quite difficult. As the D-cones are always surrounded by a crescent of strong average curvature (related to the folds that connect at the cones) that appears to be much smoother in space, detection using large values of $|L|$ is more efficient. We then apply a tracking procedure using the algorithm developed by N. Ouellette~\footnote{\url{http://web.stanford.edu/\~nto/software_tracking.shtml}}\cite{Ouellette}. In this way we obtain trajectories of the strong events. These trajectories have a finite duration as these structures have a finite lifetime and ultimately fall below the detection threshold. By differentiating the trajectories in time, we obtain their velocity. This step requires a low-pass filtering operation to remove the noise around the actual trajectories due to the errors on the estimate of the position of the structure. Determining accurately the cutoff frequency of the filter is an issue and remains somewhat arbitrary. The PDF of the velocity of the strong events is shown in fig.~\ref{pdfpiv}. Such tracking is possible only at the highest forcing intensities for which the strong events are clearly overcoming the fluctuating background due to the small scale waves. We see that the distribution of the velocity is close to Gaussian. The velocity variance seems to increase slightly with the forcing intensity and is of the order of $20$~m/s. We have no clue so far to explain this particular value for the velocity.

\begin{figure}[!htb]
\centering
\includegraphics[width=9cm]{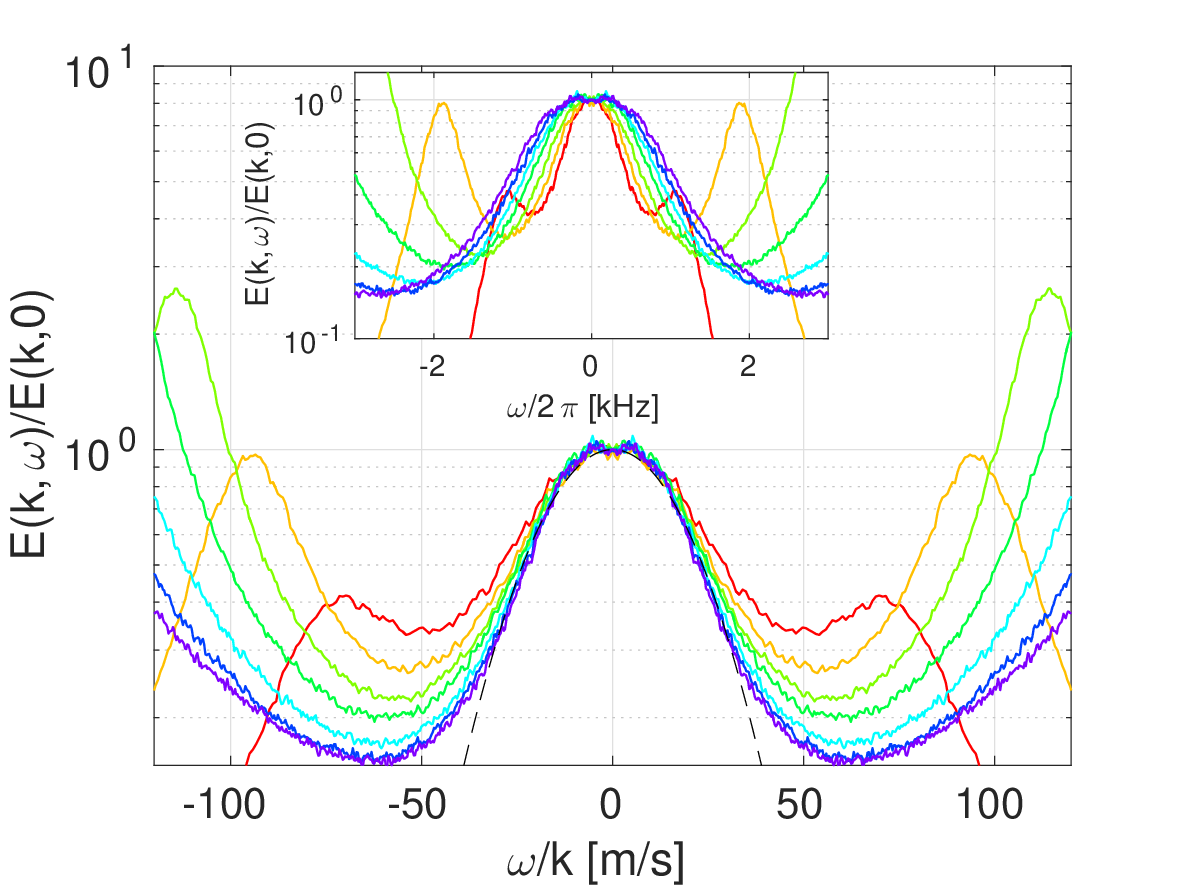}
\caption{Cuts of the $(k,\omega)$ spectrum at various values of $k/2\pi=15,20,25,30,35,40,45$~m$^{-1}$ for run 8 (strongest forcing). Main figure: $E^{\eta}(k,\omega)/E(k,0)$ as a function of $\omega/k$. The dashed line is a Gaussian shape of width 20~m/s. Insert: $E^{\eta}(k,\omega)/E(k,0)$ as a function of $\omega$}
\label{cent}
\end{figure}
Let us focus now on the central tongue (peak) of the spectrum $E^{\eta}(k,\omega)$ previously shown in figs.~\ref{stfs} and \ref{cutsp}. We consider the strongest forcing as observed in the numerical simulations in run 8 (fig.~\ref{cent}), for which the feature is the most visible. The frequency bandwidth of this peak is seen to increase with the wavenumber so that the peaks observed at various given wavenumbers do not collapse in fig.~\ref{cent}(a). By contrast, when the frequency is normalized by the wavelength $\lambda=\frac{2\pi}{k}$, the central peaks for various values of $k$ collapse fairly on each other. In this representation, the width of the central peak defines a quantity homogeneous to a velocity. The dashed line in fig.~\ref{cent}(b) is a Gaussian shape of width 20~m/s which reproduces well the shape of the peak. This figure can also be drawn for the second highest forcing and gives a similar result. At lower forcing intensities the central peak is not strong enough so that the estimate of its width is not possible with enough accuracy. 
The consistency between our spectral estimate and our direct measurement in physical space supports strongly the fact the peak observed in fig.~\ref{cent} is indeed associated to the motion of coherent structures. Furthermore, the velocity of sweeping of these structures can be estimated precisely using the spectral analysis described above. This central peak is actually similar to the space-time spectrum of isotropic homogeneous Navier-Stokes turbulence in which no waves are present and only sweeping of vortices is observed (see fig. 2 of reference~\cite{Mininni}).


\begin{table}
\centering    
\begin{tabular}{c|c|c|c|c}
run & 5 & 6 & 7 & 8\\
\hline
$u_{rms}$ 
[m/s] & 18 & 22 & 23 & 26 \\
\end{tabular}
\caption{Values of the root mean square velocities of the singularities. The major source of uncertainty on the evaluation of the velocity is the presence of noise in the trajectories. This noise has to be filtered~\cite{NMPhysD} before differentiating the trajectories to obtain the velocity. Getting the adequate cutoff frequency of the filter is not obvious in this case so that the values of the velocity given above should be taken as an order of magnitude rather than an accurate estimate.}
\label{tab-param}
\end{table}

We propose below a simple 1D model of the random motion of such structures in the spirit of previous work by Kuznetsov~\cite{Kuz}. Let us assume that some spatial structures $f(r)$ (assumed identical and isotropic for simplicity) are placed randomly in space (following a Poisson distribution) with an average density $n$. The structures translate with velocities that are distributed following a isotropic distribution. $P(u)$ is the probability distribution of the velocity $u$. Although we are not able to demonstrate it rigorously, Monte-Carlo-like simulations suggest that  the space-time spectrum of such signal may be:
\begin{equation}
E^{\eta}(k,\omega)\propto \frac{n |\hat{f}(k)|^2}{k} P(\omega/k)
\label{sps}
\end{equation}
where $\hat{f}(k)$ is the Fourier transform of a single structure $f(r)$. 

If we trust this expression, then the shape of the collapsed cuts of the spectrum (at given values of $k$) shown in fig.~\ref{cent}(b) is reminiscent of the distribution of the velocity. The shape of the observed peak is seen to follow a Gaussian variation. As has been shown above the distribution of the velocity of the structures is indeed very close to a Gaussian distribution. 

\begin{figure}[!htb]
\centering
\includegraphics[width=8cm]{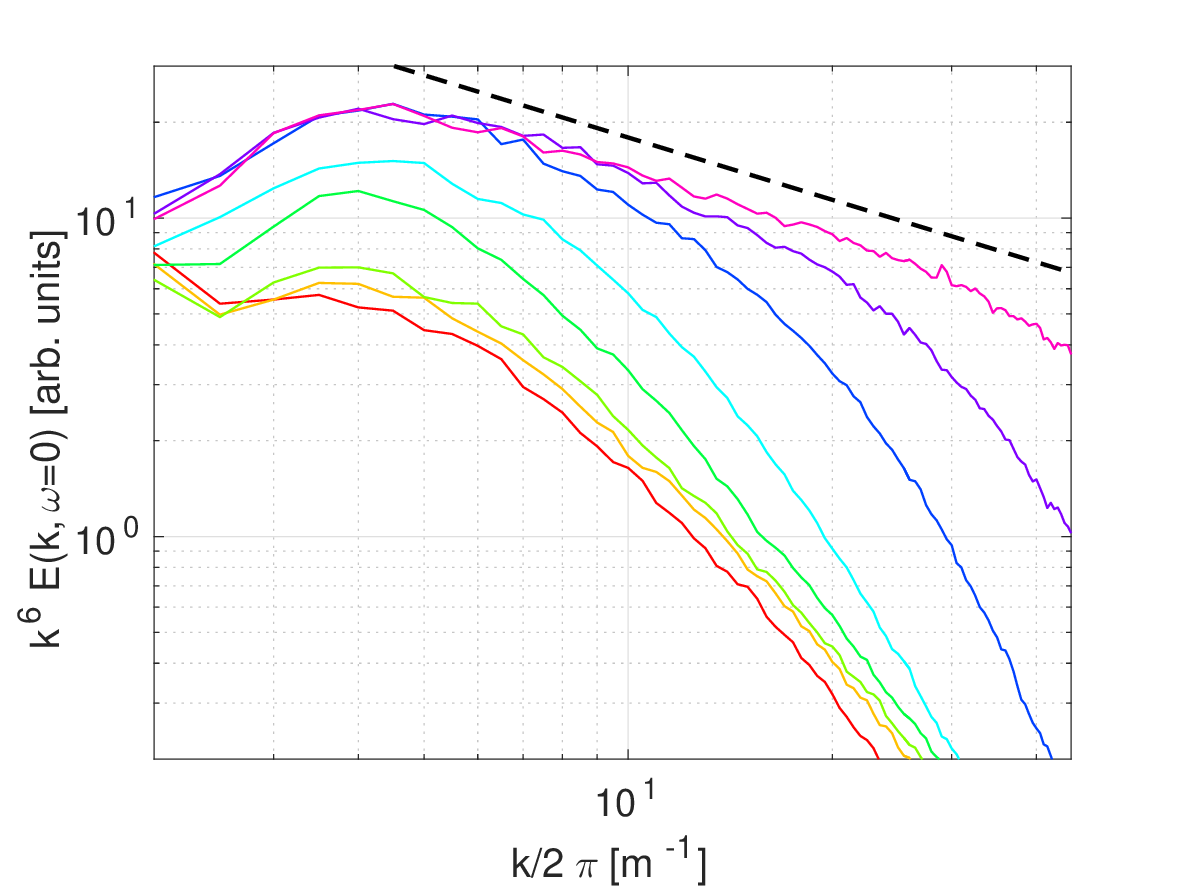}
\caption{Decay of the space-time spectrum at zero frequency $k^6E(k,\omega=0)/\iint E^{\eta}(k,\omega)dkd\omega$ for all simulations. The normalization of the spectrum means that the deformation has been normalized to be of variance 1. The dashed line shows a decay $k.^{-2/3}$. The forcing increases from bottom to top.}
\label{spw0}
\end{figure}
If one takes the case $\omega=0$ in the formula (\ref{sps}) we obtain $E(k,0)\propto \frac{n |\hat{f}(k)|^2}{k} P(0)$ which is mainly the spectrum of the structure $f(r)$. Similar spectra at $\omega=0$ for the numerical simulation are shown in fig.~\ref{spw0}. The spectrum $E(k,0)/\iint E^{\eta}(k,\omega)dkd\omega$ that is shown has been summed over directions which induces an extra $k$ factor so that the curves that are shown should correspond to $|\hat{f}(k)|^2$ following (\ref{sps}). The 3 curves corresponding to the lowest forcing are almost superimposed and correspond to cases without structures. They are not zero because of the slight spread of the energy concentrated around the dispersion relation. When the forcing is further increased the spectrum at zero frequency shows the appearance of the structures and their spectrum is then extending more and to small scales. At the strongest forcing they tend to go to a limit shape that appears to follow a power law which exponent is close to $k^{-20/3}$. This observation is slightly steeper yet comparable to the spectrum of the D-cones which has been predicted by D\"uring {\it et al.} to be proportional to $k^{-6}$ \cite{R18}. The discrepancy may be explained by the fact that the D-cones are actually 2D structures that are regularized by dissipation at small scales. Furthermore they are not isotropic at small scale and/or that they are most likely distributed in size.

\section{Statistical properties and intermittency}

\subsection{Numerical simulations}

The strong events of curvature seem to have an internal structure and a spatial organisation. These are the ingredients for statistical intermittency as was observed in Navier-Stokes turbulence~\cite{Frisch}. The statistical tools to study this intermittency are for instance the structure functions that probe the statistics at various scales. For Navier-Stokes turbulence, one often studies the spatial increments of velocity $\delta_r v=v(x+r)-v(x)$ or the gradients of the velocity gradient averaged over a sphere of radius $r$. Following a similar approach we study the evolution of the statistics of the curvature when first smoothing the picture of the deformation by low-pass filtering. In practice, the deformation field is low-pass filtered by a Gaussian kernel. The curvature of the filtered images is then computed. Note that the order of the operations of filtering and calculation of the curvature does not matter for the average curvature which is approximated by the linear operator $L_0$. By contrast, the Gaussian curvature, even evaluated using the simplified form $G_0$, is not a linear operator and the order of the calculation matters. After computing the curvatures $G$ and $L$ of the smoothed deformation with different low-pass filter cutoff scales, we compute the distributions and the moments of these curvatures. 

\begin{figure*}[!htbp]
\centering
\includegraphics[width=\textwidth]{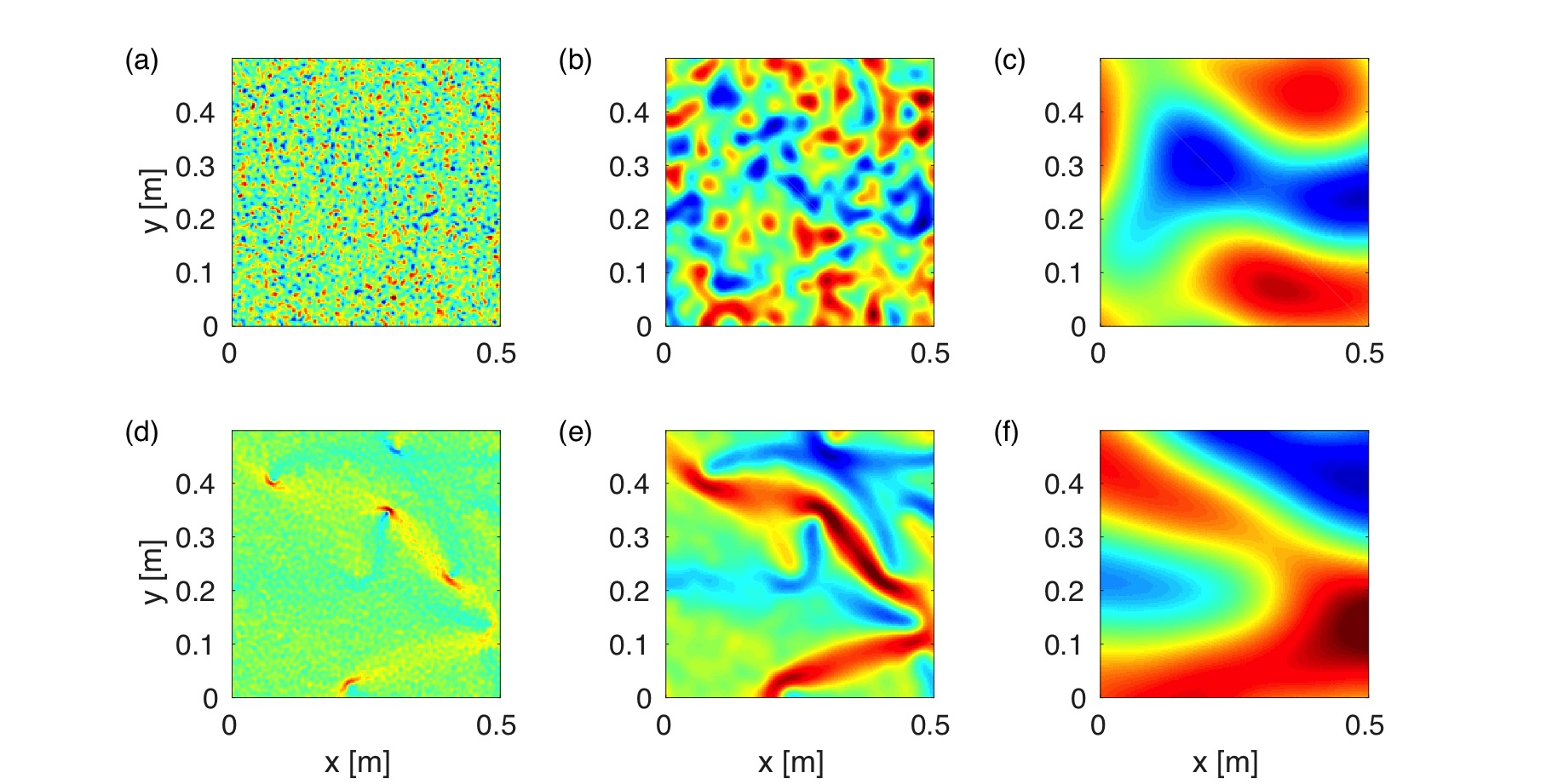}
\caption{Examples of principal curvature $L$ of the low pass filtered deformation. Top line, (a), (b), (c): weakest forcing (run~1, $P=1$). Bottom line, (d), (e), (f): strongest forcing (run~8, $P=11100$). (a) \& (d): curvature of the deformation. (b) \& (e): curvature of the deformation after smoothing over 5 pixels (1.3~cm). (c) \& (f): curvature of the deformation after smoothing over 32 pixels (8.3~cm). The images display only a portion of the full plate.}
\label{exintL}
\end{figure*}


Examples of the curvature of such low pass filtered deformation are shown in fig.~\ref{exintL}. At small forcing the picture of the curvature remains disorganized and close to Gaussian statistics regardless of the scale of smoothing. At strong forcing the intense localized structures visible at small scales are progressively filtered and filamentary organisation in the picture appears at intermediate smoothing scale. These observations are the trace of a spatial organisation of the curvature that is not self similar in scale. 

\begin{figure*}[!htb]
\centering
\includegraphics[width=\textwidth]{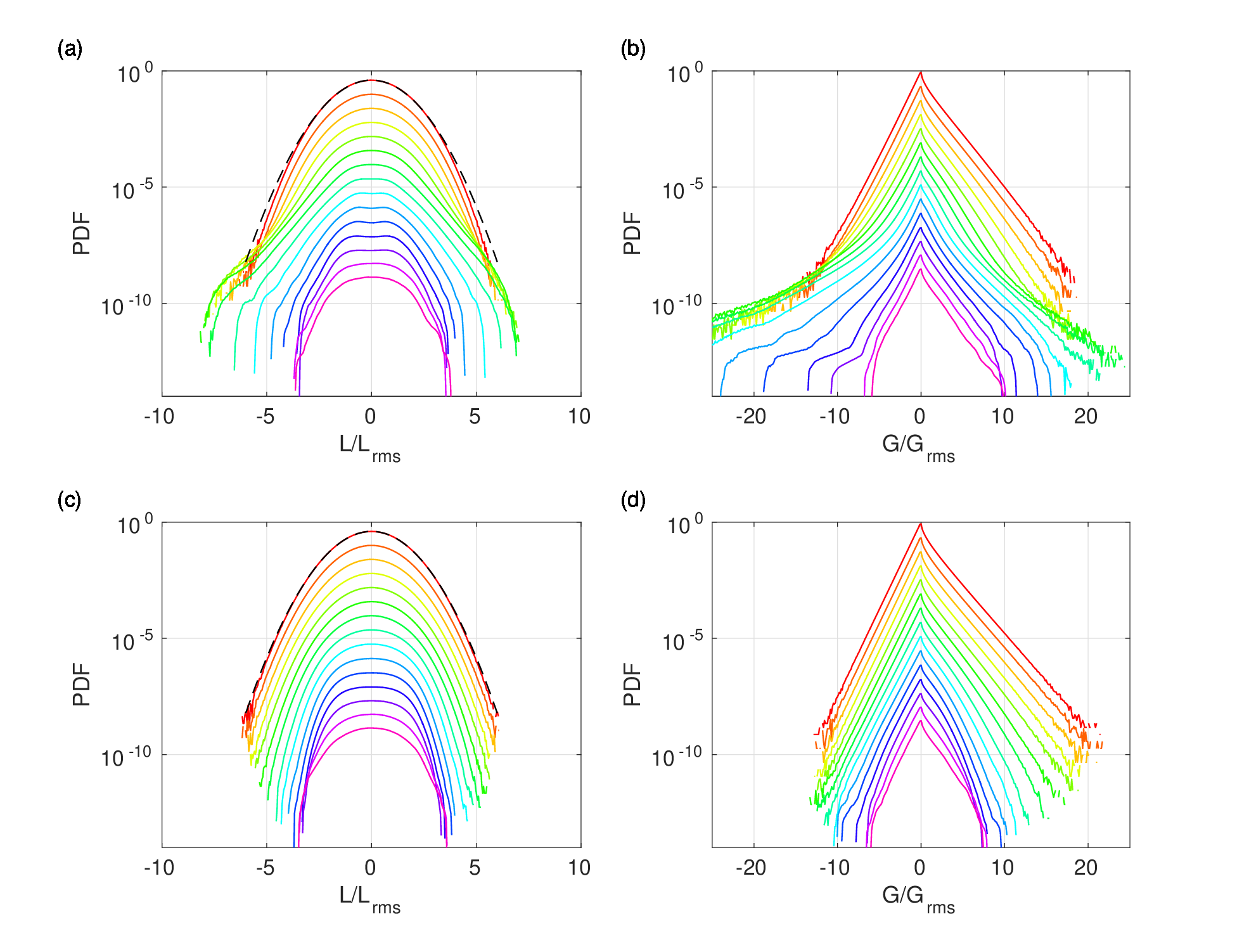}
\caption{PDF of both curvature $L$ and $G$ as a function of the smoothing  for run5 ((a) \& (b), $P=245$) and run1 ((c) \& (d), $P=1$). (a) \& (c): $L$, the dashed line shows a Gaussian distribution, (b) \& (d): $G$. In each subfigure, the lengthscale is increasing from top to bottom and the curves have been shifted vertically for clarity. The smoothing scales are from bottom to top: 0.05, 0.14, 0.29, 0.43, 0.57, 0.86, 1.1, 1.7, 2.3, 3.4, 4.6, 6.9, 9.2, 13 and 18~cm.}
\label{PDFsc1}
\end{figure*}

The lack of self similarity is visible when computing the distribution of the curvature of the smoothed deformation. In fig.~\ref{PDFsc1}, the curvatures are computed for run5 (top line). The distribution of $L$ is seen to be close to Gaussian at small scales, develops wider tails at intemediate scales and then evolves back to Gaussian at the largest scales. The distribution of $G$ also develops wider tails at intermediate scales. For run1 (weak turbulence, bottom line of fig.~\ref{PDFsc1}), the PDF are almost unchanged whatever the smoothing scale). This shows that the intermittency (understood as the lack of self similarity of the deformation field) is due to the existence of the coherent structures.

\begin{figure}[!htb]
\centering
\includegraphics[width=8.5cm]{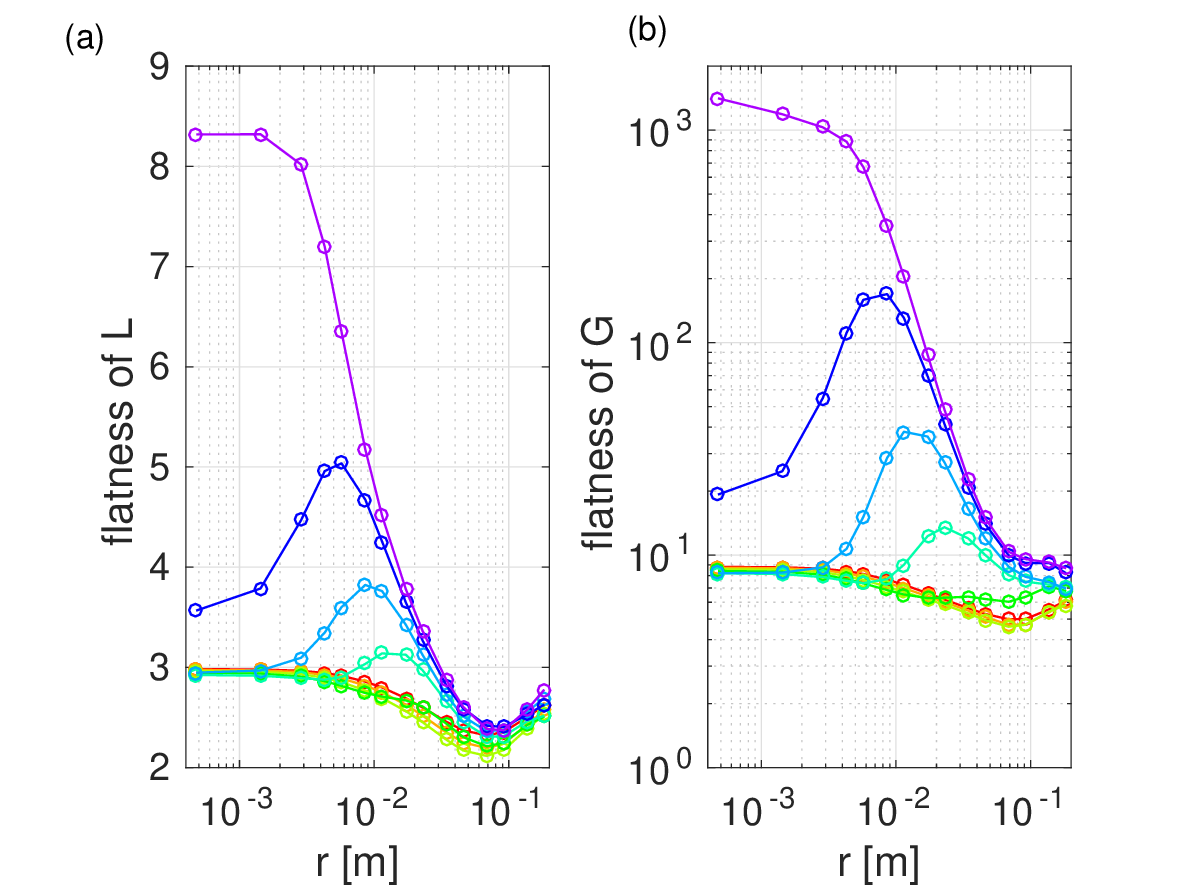}
\caption{Evolution of the flatness of (a) $L$ and (b) $G$ with scale.}
\label{flatc}
\end{figure}

A way to quantify the change of the shape of the distribution is to compute its moments made dimensionless by normalizing by the variance. For instance the flatness $F=\frac{\langle L^4\rangle}{\langle L^2\rangle^2}$ is a measure of the evolution of the width of the tails of the distribution. The evolution of the flatness with the smoothing scale is shown in fig.~\ref{flatc}. At low forcing the flatness only shows a slight evolution when increasing the scale: The average curvature evolves from 3, the value for a Gaussian random field, to about 2.5 and back to 3 at large scales. The Gaussian curvature displays a similar trend. At larger forcing intensity, the evolution of the flatness displays a maximum at intermediate scales (or order 1~cm). This shows indeed that the distribution of the PDF displays wider tails at these intermediate scales. At the stronger forcing the flatness takes very large values at small scales and decays continuously to the Gaussian value at large scale.



\subsection{Experiments}

\begin{figure}[!htb]
\centering
\includegraphics[width=9cm]{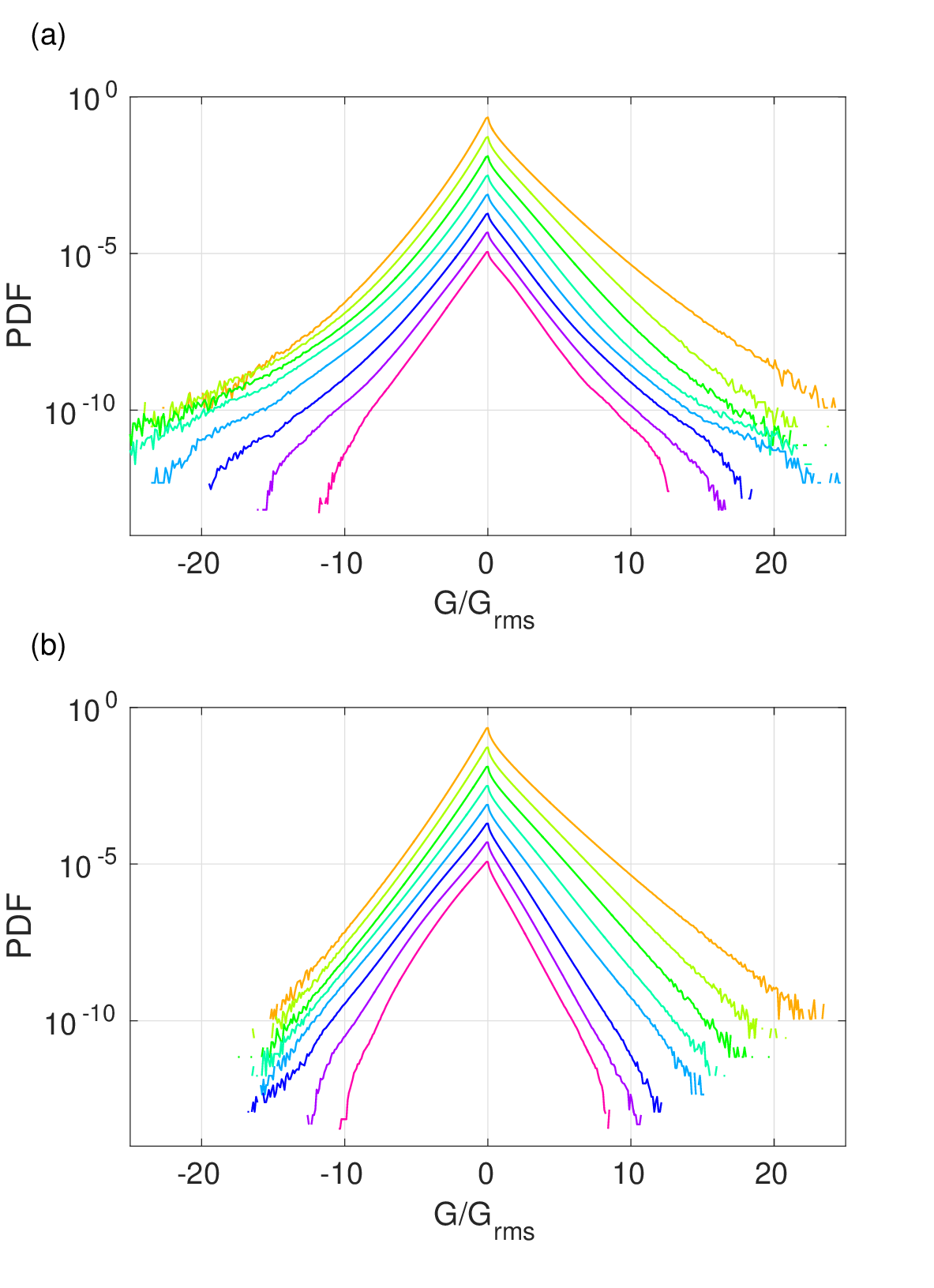}
\caption{Evolution with scale of the Gaussian curvature of the smoothed images for experiments. (a): higher forcing (runG, $P=50$). (b): smallest forcing which is not affected by the noise (runC, $P=9$).}
\label{pdfcm}
\end{figure}
The PDF of the  Gaussian curvature of the smoothed experimental deformations are shown in fig.~\ref{pdfcm} for two forcing intensities. Computing the curvature of a real image is quite challenging as it involves differentiating twice an image which contains measurement noise. Thus this cannot be achieved at the smallest forcing intensities for which the noise has a big impact. For this reason, we restrict ourselves to the analyse of a medium intensity forcing and of the largest experimental forcing. The figure~\ref{pdfcm} should be compared to the equivalent figure~\ref{PDFsc1}. A similar trend emerges: almost no evolution of the shape is observed at the smallest forcing whereas the tails of the PDF are wider at intermediate scales at the highest forcing. This widening of the tails is much less pronounced on the experiments than on the numerical simulation due to not as strongly non linear a motion, resulting itself from experimental limitations on the forcing~\cite{R23,R24}. Indeed as seen in fig.~1, the experiment with the strongest forcing is qualitatively similar to the numerical run 4. Run 4 corresponds to a mild forcing for which most of the features discussed above are not clearly visible. Nevertheless the emergence of the structures in the experiment can be observed clearly from the change of shape of the spectrum (fig.~\ref{sp}). Concerning the distribution of the curvature in fig.~\ref{pdfcm} the emergence of structure is visible only in the very far tails that correspond to extremely rare events. unfortunately we could not achieve a stronger forcing in the experiment because the instantaneous input power is strongly fluctuating. Thus the shaker must intermittently provide an instantaneous (electrical) power several orders of magnitude larger than the average, which the power supply unit can not provide despite being strongly oversized compared to the average power required (see discussion on the statistics of input power by Cadot {\it et al.}\cite{Cadot}). 

\begin{figure}[!htb]
\centering
\includegraphics[width=8.5cm]{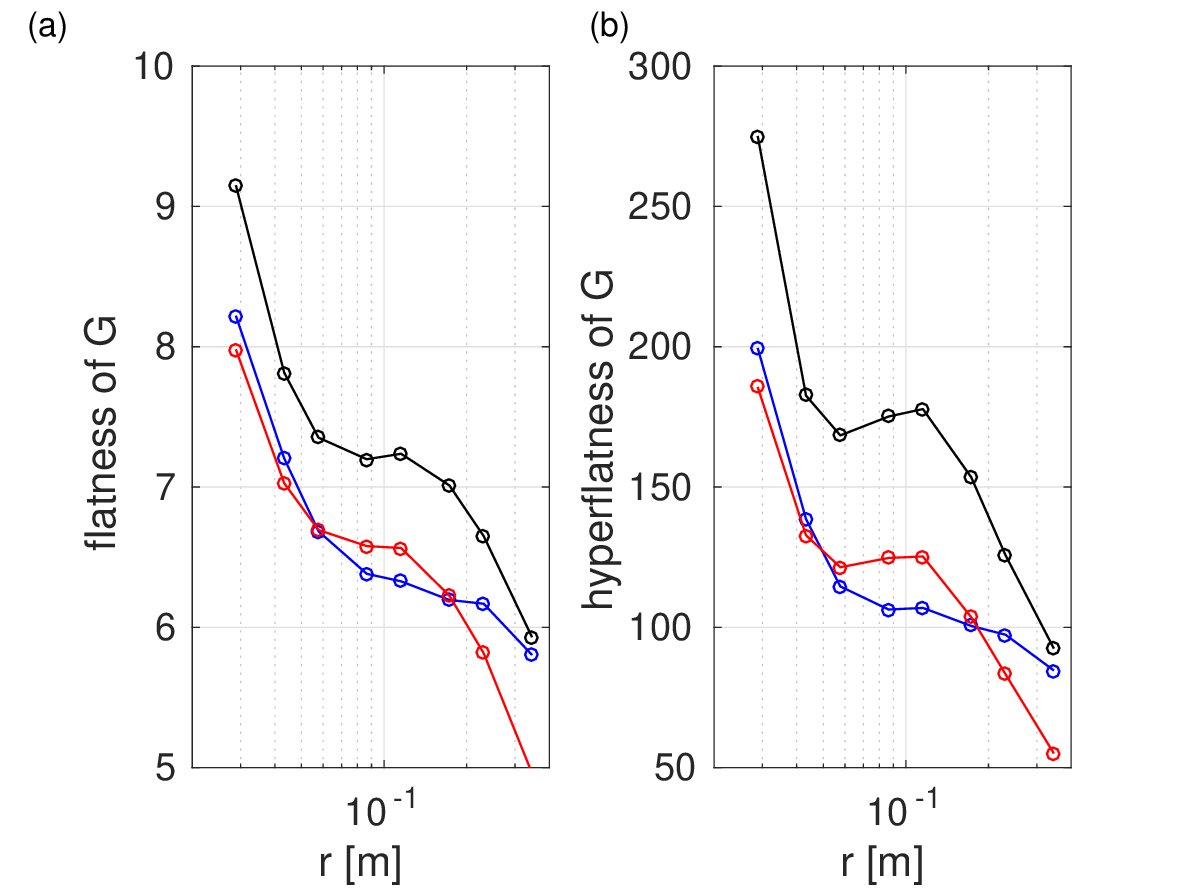}
\caption{Evolution with scale of the flatness (a) and hyperflatness (b) of the curvature for the experiment  for runs E, F, G (colors blue, red and black resp., input power $P=25$, 36 \& 50 resp.).}
\label{flatcm}
\end{figure}
This evolution of the shape of the PDF can be quantified by the flatness and hyperflatness (order 6 moment). As the change in shape of the PDF occurs in the very far tails the evolution is best seen on the hyperflatness which is a higher order moment and thus more sensitive to rare events. The change of shape of the PDF can be seen only on the statistics of the Gaussian curvature which is the only one shown. At low forcing (blue curve) the flatness and hyperflatness are quite flat except at small scales at which the noise dominates over the signal. When the forcing is increased a local maximum of the flatness (barely visible) and the hyperflatness (more visible) emerges which is in qualitative agreement with the observations for numerical simulations although not as strong (see fig.~\ref{flatc}).

\section{Concluding remarks}
We have described in this paper the gradual emergence of coherent structures in a wave turbulence system: in the case of flexural waves, coherent structures first appear at large wavelength and march toward small scales as the forcing is increased. These structures are composed of developable cones, which imprint at small scale is the crescent shape structure observed in the numerical simulation, which are connected by folds~\cite{RMPWitten}. At the smallest scales though, these structures are regularized and dominated by a bath of weakly non linear waves with small wavelength (except possibly for the highest forcing case) or by dissipation. Observe that all these structures appear, evolve dynamically and die. This network of interconnections between cones generates some intermediate scale structures, namely ridges and facets, resulting in a crumpled paper appearance. The weak turbulence scaling is confined to a range of small scales that diminishes and eventually disappears at the highest forcing. A change in the statistical properties of the system accompanies this transition: intermittency appears. 

In a thin elastic plate, we associate the occurence of intermittency with the observation of coherent structures. Indeed the apparition of intermittency at large forcing intensity is correlated with the development of wide tails of the curvature distribution at small scale, a signature of developable cones. Our observations can be compared to the case of fluid turbulence. It was suggested in the last decades that the intermittency observed in Navier-Stokes turbulence could be associated with the observed intense coherent vorticity filaments but this issue is still a matter of debate. 

The vibrating plate appears to be a relatively simple physical setup in which to observe the transition from weakly non linear wave turbulence compatible with the theory of Weak Turbulence to a regime of stronger nonlinearity that generate coherent structures. Although such behavior maybe expected generically in other wave systems the variety of possible coherent structures probably makes each system specific. In other words, this transition results from the ``breakdown of weak turbulence'' as worded by Newell {\it et al.} \cite{NNB}. The widening of the dispersion relation is related to the nonlinear time scale. \Add{The nonlinear timescale $T_{NL}(k)$ can be defined as the correlation time of the modulation of the Fourier modes as in  ref.~\cite{Miquel3}. One can write that the Fourier mode of the plate deformation as $\zeta_{\mathbf k}(t)=A_{\mathbf k}(t)e^{i\omega_kt}$ where $A_{\mathbf k}(t)$ is the modulation. The width (in frequency) of the deformation spectrum (around the dispersion relation) shown in fig.~\ref{stfs} is thus directly related to the characteristic timescale of $A_{\mathbf k}(t)$ that can be extracted from the temporal correlation of $A_{\mathbf k}(t)$~\cite{Miquel3}.} As seen in fig.~\ref{stfs}, at strong forcing the widening is so high that the two branches of the dispersion relation overlap at low frequency. This is the scenario predicted by Newell {\it et al.}: the hypothesis of scale separation between the nonlinear time scale $T_{NL}(k)$ and the linear period of the wave $T(k)$ must breakdown either at large scale or small scale depending on wether $T_{NL}/T$ is increasing of decreasing with $k$ on the KZ spectrum. For surface gravity waves the ratio $T_{NL}/T$ is increasing with $k$ so that wave turbulence becomes more and more nonlinear as the cascade proceeds to small scales and this may lead ultimately to whitecapping. By contrast, for elastic plates, the ratio $T_{NL}/T$ is decreasing when the frequency is increasing (as for capillary waves for instance). So the breakdown of weak turbulence occurs at low frequency (near the forcing scales) but the time scale separation remains valid at large enough frequency. \Add{We confirmed experimentally this breakdown in fig.~11 of~\cite{Miquel3} where we observed indeed that the nonlinear time scale is indeed comparable to the wave period at the strongest forcing (run G). Note that the coherent structures appear at large scale in fig.~\ref{ex} but are originating from stress focussing at the smallest scales}. Thus the vibrating plate corresponds to another class of systems than the case of surface gravity waves. Newell {\it et al.} suggest that for surface gravity  waves the breakdown should lead to the observation of the Philipps spectrum. This spectrum does not depend on the input power $P$ anymore as it corresponds to a saturated spectrum. Whitecap events dissipate extra energy so that to remain on the Philipps spectrum. This is not what is observed in the ideal elastic plate. The spectrum changes from being proportional to $P^{1/3}$ in the weak case to being proportional to $P$ in the strong case. This corresponds to the behavior predicted by Kuznetsov~\cite{Kuz} for the spectrum of point singularities that is expected to be proportional to $P$. Increasing the forcing generates more and more singularities. It should be noted though that dissipation may alter this scenario in real plates. In our simulations, the dissipation is simply linear in the velocity and only present at small scales so that strongly nonlinear dissipation analogous to that occurring in whitecap events of sea waves cannot exist. We also assume the F\"oppl-von Karman equation to remain valid, as a toy model of weak turbulence, rather than to simulate real plates. In actual plates, it is likely that, at extremely strong forcing, strongly non linear deformation can lead to nonlinear dissipation and even irreversible events due to plasticity. It may be possible that for plates that would be large enough another transition could occurs from the Kuznetsov-like regime to a saturated Philipps-like spectrum at large scales and thus a spectrum that becomes independent of $P$ but this remains very speculative and clearly beyond the possibilities of our simulations where the linear dissipation is enough to regularize the singularities at small scale.

\section{aknowledgements}
This project has received funding from the European Research Council (ERC) under the European Union's Horizon 2020 research and innovation programme (grant agreement No 647018-WATU). NM was supported by Institut Universitaire de France. We thank O. De~Marchi, C. Bonamy and G. Moreau for their help to run the numerical simulations.
\bibliography{plaquexii.bib}

 \end{document}